   \newcommand{\be}[0]{\begin{equation}}
   \newcommand{\ee}[0]{\end{equation}}
   \newcommand{\ba}[0]{\begin{eqnarray}}
   \newcommand{\ea}[0]{\end{eqnarray}}
\newcommand{\MSb}{\overline{MS}}

\newcommand{\LMSb}{\Lambda_{\MSb}}
\newcommand{\alphazero}{\overline{\alpha}_0}

\newcommand{\reffg}[1]{Fig.\ref{fg:#1}}
\documentclass[12pt]{article}
\usepackage{epsfig} \usepackage{amssymb} \usepackage{amsfonts}
\usepackage{epsf,psfrag}
\begin{document}
\Large
\hfill\vbox{\hbox{IPPP/04/47}
            \hbox{DCPT/04/94}}

\nopagebreak

\vspace{0.75cm}
\begin{center}
\LARGE
{\bf ${\Lambda}$-based QCD perturbation theory-: side-stepping the scheme dependence problem}
\vspace{0.6cm}
\Large

M.~J.~Dinsdale\footnote{email:{\tt m.j.dinsdale@durham.ac.uk}} and C.~J.~Maxwell\footnote{email:{\tt c.j.maxwell@durham.ac.uk}}

\vspace{0.4cm}
\large
\begin{em}
Institute for Particle Physics Phenomenology, University of Durham,
South Road, Durham, DH1 3LE, UK
\end{em}

\vspace{1.7cm}

\end{center}
\normalsize
\vspace{0.45cm}

\centerline{\bf Abstract}
\vspace{0.3cm}
%%%%%%%%%%%%ABSTRACT%%%%%%%%%%%%%%%%%%
We advocate the replacement of standard ${\alpha}_{s}(\mu)$-based QCD
perturbation theory, in which the coupling and truncated perturbative
predictions are dependent on the chosen renormalisation scheme, by a
${\Lambda}$-based approach in which QCD observables are directly related
to the dimensional transmutation parameter, ${\Lambda}$, of the theory.
The shortcomings of the standard approach are emphasised by formulating it
in the ${\Lambda}$-based language. We show how the ${\Lambda}$-based approach
can be extended to include massive quarks. We also consider the use of ${\Lambda}$-based perturbation theory
in the case when a resummation of large infrared logarithms is required,
performing a phenomenological analysis for the ${e}^{+}{e}^{-}$ event shape
observables thrust and heavy jet mass. We show by fitting to
data that smaller power corrections are required than in the standard approach.

\newpage
\section*{1 Introduction}
The process of removing ultraviolet divergences in defining the renormalised
coupling of quantum field theories is not uniquely specified since the finite
part may be adjusted at will. This renormalisation scheme (RS) dependence means that
renormalisation group improved (RG-improved) fixed-order perturbative
predictions depend on the arbitrary renormalisation convention used. The small value
of the coupling in Quantum Electrodynamics means that this is not a practical
problem. As is well-known, perturbative calculations with the conventionally used
on-shell renormalisation scheme are in exceptionally good agreement with experimental
measurements of, for instance, the anomalous magnetic moment of the electron. The
situation in Quantum Chromodynamics is not so favourable and the scheme dependence
problem is a source of serious ambiguity in making precise perturbative QCD
predictions, for a review see \cite{r1}. We shall refer to the standard approach
as ``${\alpha}_{s}(\mu)$-based'' perturbation theory (${\alpha}_{s}(\mu)$BPT). The renormalisation scale
is conventionally chosen to be ${\mu}=xQ$, where $Q$ is the physical energy scale
of the process, for instance the centre of mass energy $\sqrt{s}$ in ${e}^{+}{e}^{-}$
annihilation. The modified minimal subtraction ($\overline{MS}$) procedure \cite{r2}
is most commonly used where the ${\ln}(4\pi)-{\gamma}_{E}$ which always arises
using dimensional regularization is subtracted off.
Both the choice of the dimensionless
number $x$ in the scale choice and the subtraction procedure are of course
completely arbitrary. Worryingly for many QCD observables calculated perturbatively
at next-to-leading order (NLO) in perturbation theory, the unphysical $x$-dependence
of the perturbative prediction is quite strong. The conventional attitude to
this is to vary $x$ around the so-called ``physical scale'' $\mu=Q$
, i.e. $x=1$, between say $x=1/2$ and $x=2$, the resulting ${\alpha}_{s}(xQ)$ values
are then evolved to $Q={M}_{Z}$ using the QCD beta-function equation truncated at
the appropriate order, and a central value of ${\alpha}_{s}(M_Z)$ corresponding
to $x=1$ together with a ``theoretical error bar'' corresponding to the variation
of $x$ over the chosen range, are obtained. The difficulty with this approach is that the
range of scales to be considered is completely {\it ad hoc}. There is nothing
sacrosanct about the choice $\mu=Q$. Indeed the meaning of $\mu$ depends on
the subtraction procedure employed. For instance the choice $\mu=Q$ with
$\overline{MS}$ subtraction gives exactly the same result as $\mu=0.38Q$ with
$MS$ subtraction where the $\ln{4}{\pi}-{\gamma}_{E}$ is not subtracted off.
Clearly the choice of scale cannot be physically motivated. Unfortunately
there are many arguments in the literature which attempt to do just this.
A recent example is Higgs production by bottom quark fusion where a kinematical
argument is used to infer the correct choice of the factorization scale \cite{r3}.
For truncated fixed-order perturbative predictions it was shown in \cite{r4} how
the RS-dependence of the renormalised coupling and the perturbative cofficients
can be parametrised by the combination ${\mu}/{\Lambda}_{RS}$, and by the
non-universal coefficients of the beta-function. Here ${\Lambda}_{RS}$ corresponds
to the dimensional transmutation parameter with a particular choice of
subtraction procedure for removing UV divergences. The beta-function is conventionally
taken to be truncated at the same order as the perturbation series. Whilst the
truncated results depend on RS, the formal all-orders sum of the perturbation
series does not. It was therefore advocated in Ref.\cite{r4} that the RS in
each order should be chosen according to the Principle of Minimal Sensitivity (PMS),
that is so that the approximant is stationary with respect to variation of
the scheme parmeters. Whilst this idea can be motivated by a number of persuasive
toy examples one cannot make any firm statements about the goodness of
approximation of the PMS fixed-order results. One is making the best of the
RS-dependence problem with which standard fixed-order RG-improved ${\alpha}_{s}(\mu)$BPT is afflicted.\\

In this paper we wish to propose an alternative ${\Lambda}$-based formulation of perturbative
QCD ($\Lambda$BPT). The idea is extremely simple, and follows from dimensional analysis.
It is very closely related to the discussion of \cite{r5}, and is essentially the method of effective charges
introduced by Grunberg \cite{r6,r7}, although as we shall explain, ${\Lambda}$BPT can, at least in principle,
be defined even if the QCD observable is {\it not} an effective charge. For a dimensionless
QCD observable ${\cal{R}}(Q)$ depending on the energy scale $Q$, and with massive quarks with pole masses
$\{{m}_{j}\}$
one will have on general dimensional analysis grounds,
\be
{\cal{R}}(Q)={\Phi}\left(\frac{\Lambda}{Q},\frac{\{{m}_{j}\}}{Q}\right)\;.
\ee
Here $\Lambda$ is a dimensionful constant related to the dimensional transmutation
parameter of the theory. We can invert ${\Phi}$ to obtain,
\be
\frac{\Lambda}{Q}={\tilde{\Phi}}\left({\cal{R}}(Q),\frac{\{{m}_{j}\}}{Q}\right)\;,
\ee
and so the ${\Lambda}$ parameter can be directly related to the measured observable.
The key point to realise is that the dimensional transmutation parameter which
is associated directly with broken scale invariance is the fundamental parameter. The
renormalisation prescription used to remove UV divergences will determine the
value of $\Lambda$. However ${\Lambda}$'s corresponding to different prescriptions
can be {\it exactly} related given a one-loop perturbative calculation in the
two schemes \cite{r8}. Since $\overline{MS}$ subtraction is most commonly employed ${\Lambda}_{\overline{MS}}$
can serve as the fundamental QCD parameter. Clearly the function ${\tilde{\Phi}}({\cal{R}}(Q),{\{}{m}_{j}{\}}/Q)$
which determines ${\Lambda}_{\overline{MS}}$ does not involve the RS parameters
on which the standard ${\alpha}_{s}(\mu)$-based fixed-order results depend, and
as reconstructed from the perturbative information it will automatically involve
RS-invariant combinations of perturbative coefficients and beta-function coefficients,
and will be completely independent of the renormalisation scale $\mu$. The explicit form
of the function ${\tilde{\Phi}}$ is easily obtained in the case that ${\cal{R}}(Q)$ is
simply related to an effective charge \cite{r9}. An effective charge corresponds
to an observable having a perturbative expansion,
\be
{\cal{R}}(Q)=a(\mu)+\sum_{n>0}{r_n}{a(\mu)}^{n+1}\;,
\ee
where $a(\mu)\equiv{\alpha}_{s}(\mu)/{\pi}$ is the RG-improved coupling.
 The coupling $a(\mu)$ will satisfy
the beta-function equation,
\be
\frac{d{a}}{d{\ln}\mu}={\beta}(a)=
-b{a}^{2}\left(1+ca+\sum_{n>1}{c}_{n}{a}^{n}\right)\;,
\ee
where $b=(33-2{N}_{f})/6$ and $c=(153-19{N}_{f})/12b$ are the universal beta-function
coefficients for $N_f$ active quark flavours. If we assume massless quarks, so that
${m}_{i}=0$, then dimensional analysis implies that,
\be
\frac{d{\cal{R}}(Q)}{dQ}=\frac{{\rho}({\cal{R}}(Q)}{Q}\;,
\ee
where ${\rho}({\cal{R}}(Q))$ is a dimensionless function of ${\cal{R}}$, which can be
recast as,
\be
\frac{d{\cal{R}}}{d{\ln}Q}={\rho}({\cal{R}}(Q))\;.
\ee
The function ${\rho}({\cal{R}}(Q)$ may be obtained perturbatively by the following
algebraic steps. Starting from the perturbation series in Eq.(3) one chooses
$\mu=Q$. Differentiating with respect to $\ln{Q}$ term-by-term and using the
beta-function equation of Eq.(4) one then obtains ${d{\cal{R}}}/d{{\ln}Q}$ as a power
series in $a(Q)$. Finally, one can invert the perturbation series in Eq.(3)
to obtain $a({\cal{R}})$ as a power series in ${\cal{R}}$. In this way one
finds,
\be
{\rho}({\cal{R}})=-b{{\cal{R}}}^{2}\left(1+c{\cal{R}}+\sum_{n>1}{\rho}_{n}{{\cal{R}}}^{n}\right)\;,
\ee
where the ${\rho}_{n}$ are RS-invariant and $Q$-independent combinations of
the perturbative coefficients $r_i$ and beta-function coefficients $c_i$,
\ba
\label{eq:rho_n}
{\rho}_{2}&=&{c}_{2}+{r}_{2}-c{r}_{1}-{r}_{1}^{2}
\nonumber \\
{\rho}_{3}&=&{c}_{3}+2{r}_{3}-4{r}_{1}{r}_{2}-2{r}_{1}{\rho}_{2}-c{r}_{1}^{2}+2{r}_{1}^{3}\;.
\\
\vdots
\nonumber
\ea
As we shall review in Section 2 one can then integrate Eq.(6), and the ${\Lambda}$ parameter
arises as a constant of integration. One finally obtains the relation \cite{r9},
\be
{\Lambda}_{\cal{R}}=Q{\cal{F}}({\cal{R}}(Q)){\cal{G}}({\cal{R}}(Q))\;.
\ee
The integration contstant ${\Lambda}_{\cal{R}}$ is exactly related to ${\tilde{\Lambda}}_{\overline{MS}}$
(the tilde implying the $\Lambda$ definition favoured in Ref.\cite{r4}) with
\be
{\Lambda}_{\cal{R}}={e}^{r/b}{\tilde{\Lambda}}_{\overline{MS}}\;,
\ee
Here $r$ denotes the NLO perturbative coefficient $r_1$ of Eq.(3) in the $\overline{MS}$ scheme
with renormalisation scale ${\mu}=Q$. ${\cal{F}}({\cal{R}})$ is the {\it universal} function,
\be
{\cal{F}}({\cal{R}})={e}^{-1/b{\cal{R}}}{(1+1/c{\cal{R}})}^{c/b}\;,
\ee
and the function ${\cal{G}}({\cal{R}})$ depends on the observable and has the
form,
\be
{\cal{G}}({\cal{R}})=1+\sum_{n=1}^{\infty}{g}_{n}{{\cal{R}}}^{n}\;.
\ee
The ${g}_{n}$ coefficients are RS-invariant and $Q$-independent combinations
of the ${\rho}_{n}$ in Eq.(8). The first few being,
\ba
{g}_{1}&=&-\frac{{\rho}_{2}}{b}
\nonumber \\
{g}_{2}&=&-\frac{{\rho}_{3}}{b}+\frac{{\rho}_{2}^{2}}{2{b}^{2}}-c\frac{{\rho}_{2}}{b}\;.
\\
\vdots
\nonumber
\ea
Eqs. (9), (10) then enable direct extraction of ${\Lambda}_{\overline{MS}}$ from the
measured data for ${\cal{R}}$. Such direct extractions of ${\Lambda}_{\overline{MS}}$ were
carried out for data on ${e}^{+}{e}^{-}$ annihilation event shape observables in \cite{r9}.
 The approach is very closely related to the method of effective charges
of Grunberg \cite{r7}, and also to the Renormalisation Scheme Invariant
Perturbation Theory (RESIPE) proposal of \cite{r10}, and the Complete Renormalisation
Group Improvement (CORGI) approach of \cite{r11}. The purpose of introducing the
$\Lambda$BPT terminology is to emphasise the fundamental importance of the dimensional
transmutation parameter $\Lambda$, and the dimensional analysis from which it
follows. It should be noticed that dimensional analysis implies that an equation
of the form of Eq.(6) holds even if ${\cal{R}}(Q)$ is {\it not} an effective charge.
However, it is not then easy to explicitly integrate Eq.(6). Hence, for instance, it is not
known how to arrive at a result of the form of Eq.(9) relating ${\cal{R}}$ and
$\Lambda$ if ${\cal{R}}$ is an observable involving the convolution of structure
functions and hard scattering partonic cross-sections. As we shall review in Section 2
structure function moments can, however, be directly related to effective charges.

The plan of this paper is to begin in Sec.2 by reviewing ${\Lambda}$BPT for effective charges,
as discussed in Ref.\cite{r9}. In Sec. 3 we show how the standard fixed-order
${\alpha}_{s}(\mu)$BPT can be written in the $\Lambda$BPT
language, this in turn will be used to emphasise that the standard approach is
very likely to provide an erroneous determination of $\Lambda$. In Sec.4 we show
how to generalise $\Lambda$BPT to the case of massive quarks, in which case Eq.(6)
is no longer a {\it separable} differential equation. Nevertheless we show that
for mass-independent renormalisation schemes such as ${\overline{MS}}$ Eq.(9)
still holds relating ${\cal{R}}$ and $\Lambda$, but where now the ${\rho}_i$
invariants of Eqs.(8) are $Q$-dependent. In Sec. 5 we then show how to generalise
$\Lambda$BPT to include a resummation of large infrared logarithms in the ${e}^{+}{e}^{-}$ annihilation
event shape observables thrust and heavy jet mass. We find that significantly smaller
power corrections are needed to fit the data than with the standard resummations based
on the physical choice of renormalisation scale. This extends the recent analysis
of the DELPHI collaboration \cite{r12} where $\Lambda$BPT was applied to the analysis of event
shape {\it means}, and where it was found to be possible to obtain consistent fits
to data without any power corrections. We consider the distributions themselves. Finally, Sec. 6
contains a Discussion and Conclusions.

\section*{2 $\Lambda$BPT for effective charges}
The dimensional analysis result of Eq.(6) is in the form of a separable differential
equation and may be straightforwardly integrated to obtain
\be
\ln\frac{Q}{{\Lambda}_{\cal{R}}}=\int_{0}^{{\cal{R}}(Q)}\frac{dx}{{\rho}(x)}+{\kappa}\;.
\ee
The constant of integration which arises is necessarily {\it infinite} to ensure that
asymptotic freedom, ${\cal{R}}(\infty)=0$, holds. It can be split into $\ln{\Lambda}_{\cal{R}}+\kappa$,
where ${\Lambda}_{\cal{R}}$ is a finite, dimensionful and observable-dependent scale, and $\kappa$
is a {\it universal} infinite constant. The integral in Eq.(14) diverges at $x=0$ since from Eq.(7)
$dx/{\rho}(x)=1/[{x}^{2}(1+cx+\ldots)]$. It is then clear that to cancel this infinity the infinite part
of the integration constant $\kappa$ must be of the form,
\be
\kappa=-\int_{0}^{C}\frac{dx}{-b{x}^{2}(1+cx+{\Delta}(x))}\;,
\ee
where ${\Delta}(x)/{x}^{2}$ is required to be finite at $x=0$, but is otherwise
arbitrary, and $C$ is an arbitrary upper limit of integration. The choices ${\Delta}(x)=0$
and $C=\infty$ are convenient, but different choices can be completely absorbed by a suitable
redefinition of ${\Lambda}_{\cal{R}}$. Defining $\kappa$ in this way we can regroup Eq.(14) in
the form
\be
\label{eq:blogQfromFG}
b\ln\frac{Q}{{\Lambda}_{\cal{R}}}=\int_{{\cal{R}}(Q)}^{\infty}\frac{dx}{{x}^{2}(1+cx)}+
\int_{0}^{{\cal{R}}(Q)}{dx}\left[\frac{b}{{\rho}(x)}+\frac{1}{{x}^{2}(1+cx)}\right]\;.
\ee
Denoting the integrals on the r.h.s. of Eq.(16) by $F({\cal{R}})$ and $G({\cal{R}})$ respectively one finds
\be
F({\cal{R}})=\frac{1}{{\cal{R}}}+c{\ln}\left[\frac{c{\cal{R}}}{1+c{\cal{R}}}\right]\;.
\ee
Exponentiating yields Eq.(9), with ${\cal{F}}({\cal{R}})$ and ${\cal{G}}({\cal{R}})$ as
defined in Eqs.(11), (12). The coefficients ${g}_k$ in Eq.(13) follow on expanding
${\exp}[-G({\cal{R}})/b]$ as a power series in ${\cal{R}}$. We finally need to relate
${\Lambda}_{\cal{R}}$ which is observable-dependent, to the universal ${\tilde{\Lambda}}_{\overline{MS}}$.
This is easily done since on rearranging Eq.(9) and taking the limit as $Q\rightarrow\infty$ one finds
\be
{\Lambda}_{\cal{R}}=\lim_{Q\rightarrow{\infty}}Q{\exp}(-F({\cal{R}}(Q))/b)\;,
\ee
where the fact that $\lim_{Q\rightarrow\infty}G({\cal{R}}(Q))=G(0)=0$ has been used.
If ${\tilde{\Lambda}}_{\overline{MS}}$ is obtained by integrating up the ${\overline{MS}}$
beta-function equation for $a(Q)$, the $\overline{MS}$ coupling with $\mu=Q$,
\be
\frac{d{a}(Q)}{d{\ln}Q}={\beta}_{\overline{MS}}(a)\;,
\ee
with the same conventions for $\kappa$, then one similarly finds
\be
{\tilde{\Lambda}}_{\overline{MS}}=\lim_{Q\rightarrow{\infty}}Q{\exp}{(-F(a(Q))/b)}\;.
\ee
Noting that
\be
F({\cal{R}})\sim{F(a(Q))}-r+\ldots\;,
\ee
where the ellipsis denotes terms which vanish as $Q\rightarrow\infty$, one can then read off
the exact relation
\be
{\Lambda}_{\cal{R}}={e}^{r/b}{\tilde{\Lambda}}_{\overline{MS}}\;.
\ee
The tilde over $\Lambda$ reflects the fact that the infinite constant $\kappa$
has been defined with the convention adopted in \cite{r4}, whereas the
commonly assumed convention \cite{r13} corresponds to $\kappa\rightarrow\kappa-cln(b/2c)$,
one then has
\be
{\Lambda}_{\overline{MS}}={\left(\frac{2c}{b}\right)}^{c/b}{\tilde{\Lambda}}_{\overline{MS}}\;.
\ee
We can then arrive at the desired relation between ${\Lambda}_{\overline{MS}}$ and ${\cal{R}}$,
\be
\label{eq:LambdaExtractor}
{\Lambda}_{\overline{MS}}=Q{\cal{K}}_{\cal{R}}^{\overline{MS}}{\cal{F}}({\cal{R}}(Q)){\cal{G}}({\cal{R}}(Q))\;,
\ee
where we have defined the observable-dependent normalisation constant
\be
{\cal{K}}_{\cal{R}}^{\overline{MS}}\equiv{e}^{-r/b}{(2c/b)}^{c/b}\;.
\ee
Notice that ${\cal{K}}_{\cal{R}}^{RS}$ is the only part of the expression dependent on
the subtraction convention used to remove ultraviolet divergences. Changes in
subtraction convention are trivial in that ${\cal{K}}_{\cal{R}}$ is simply scaled
by a constant precisely calculable given that the NLO coefficients $r$ are known
in the two RS's \cite{r8}. We have for instance
\be
{\cal{K}}_{\cal{R}}^{MS}={\exp}\left(\frac{{\gamma}_{E}-{\ln}(4\pi)}{2}\right)
{\cal{K}}_{\cal{R}}^{\overline{MS}}\;.
\ee
Without a NLO calculation the normalisation factor ${\cal{K}}_{\cal{R}}$ is
unknown and so tree-level calculations by themselves cannot be used to determine
${\Lambda}_{\overline{MS}}$, equivalently in the standard ${\alpha}_{s}(\mu)$BPT
approach tree-level calculations have a monotonic $\mu$ dependence and are hopelessly
ambiguous. Given a NLO calculation ${\cal{K}}_{\cal{R}}$ is known, but the coefficients
${g}_i$ in Eq.(13) are unknown without a ${\rm{N}}^{i+1}$LO calculation, so
that the state of our knowledge is ${\cal{G}}({\cal{R}})=1$, and the deviation from unity
is simply unknown without a NNLO calculation, which would give the estimate
\be
{\cal{G}}({\cal{R}})=1+{g}_{1}{\cal{R}}=1-\frac{{\rho}_{2}}{b}{\cal{R}}\;,
\ee
with ${\rho}_{2}$ the NNLO RS invariant defined in Eqs.(8).
As we shall discuss in the next Section the standard ${\alpha}_{s}(\mu)$BPT
is equivalent to the $\Lambda$BPT with RS-dependent ${g}_{i}$ coefficients, and
we shall argue that it is likely to give misleading determinations of ${\Lambda}$.\\

We finally note that although moments of DIS structure functions apparently involve
a factorisation scale and a renormalisation scale, they can be directly related
to effective charges \cite{r7} and ${\Lambda}$BPT applied \cite{r11}. The
${n}^{\rm{th}}$ moment of a non-singlet structure function $F(x)$,
\be
{\cal{M}}_{n}(Q)=\int_{0}^{1}{x}^{n-2}F(x){dx}\;,
\ee
can be factorised into an operator matrix element and coefficient function
\be
{\cal{M}}_{n}(Q)=<{\cal{O}}_{n}(M)>{\cal{C}}_{n}(Q,a(\mu),\mu,M)\;.
\ee
Here $M$ is an arbitrary factorisation scale, and $\mu$ the renormalisation scale.
The operator ${\cal{O}}_{n}$ will have a corresponding anomalous dimension ${\gamma}_{{\cal{O}}_{n}}(a)$
such that
\be
\frac{M}{<{\cal{O}}_{n}>}\frac{\partial{\cal{O}}_{n}}{\partial{M}}={\gamma}_{{\cal{O}}_{n}}=
\sum_{{k}\ge{0}}-{\gamma}^{(n)}_{k}{a}^{k+1}\;,
\ee
where $a$ denotes $a(M)$.
The first anomalous dimension coefficient ${\gamma}^{(n)}_{0}$ is independent of the factorisation
scheme (FS), but the higher coefficients are FS-dependent. The coefficient function has the
perturbative expansion
\be
{\cal{C}}_{n}=1+\sum_{k\ge{1}}{r}_{k}^{(n)}{\tilde{a}}^{k}\;,
\ee
where $\tilde{a}$ denotes $a(\mu)$. Integrating Eq.(30) one obtains
\be
{\cal{M}}_{n}={A}_{n}{\left(\frac{ca}{1+ca}\right)}^{{\gamma}^{(n)}_{0}/b}{\exp}({\cal{I}}(a))(1+\sum_{k\ge{1}}{r}_{k}^{(n)}{\tilde{a}}^{k})\;.
\ee
$A_n$ is an entirely non-perturbative normalisation constant. If one expands $a(M)$ as a series
in $a(\mu)$ one can recast Eq.(32) in the form
\be
{\cal{M}}_{n}={A}_{n}{[{c{\hat{\cal{R}}}}_{n}]}^{{\gamma}_{0}^{(n)}/b}\;,
\ee
where ${\hat{\cal{R}}}_{n}$ is an effective charge, with the perturbative expansion
\be
{\hat{\cal{R}}}_{n}=a+\sum_{k\ge{1}}{\hat{r}}_{k}^{(n)}{a}^{k+1}.
\ee
The ${\hat{r}}_{k}^{(n)}$ depend on $\mu$ and the non-universal beta-function
coefficients, but are $M$-independent, for instance
\be
{\hat{r}}_{1}^{(n)}=b{\ln}\left(\frac{\mu}{{\tilde{\Lambda}}_{\overline{MS}}}\right)
-b{\ln}\left(\frac{M}{{\tilde{\Lambda}}_{\overline{MS}}}\right)
-\frac{b}{{\gamma}_{0}^{(n)}}{r}_{1}^{(n)}
+\frac{{\gamma}_{1}^{(n)}}{{\gamma}_{0}^{(n)}}-c\;.
\ee
Thus ${\Lambda}$BPT applies and ${\hat{\cal{R}}}_{n}={({\cal{M}}_{n}/{A}_{n})}^{b/{\gamma}_{0}^{(n)}}/c$ is related to ${\Lambda}_{\overline{MS}}$
by Eq.(24), where $r$ in ${\cal{K}}_{\cal{R}}^{\overline{MS}}$ is the NLO coefficient ${\hat{r}}_{1}^{(n)}$
of Eq.(34) with $\mu=Q$. Simultaneous fits for $A_n$ and ${\Lambda}_{\overline{MS}}$ have
been performed on CCFR data for $F_3$ neutrinoproduction DIS moments in Ref.\cite{r14}.

\section*{3 Standard RG-improvement in ${\Lambda}$BPT language}
In this section we wish to clarify the sense in which we claim that ${\Lambda}$BPT
side-steps the scheme dependence problem inherent in the standard ${\alpha}_{s}(\mu)$BPT
approach. We shall show that the standard approach corresponds precisely to ${\Lambda}$BPT
with RS-dependent ${g}_i$ coefficients. Suppose that we have completed a
${\rm{N}}^n$LO calculation of ${\cal{R}}$, then we can split ${\cal{G}}({\cal{R}})$ in
Eq.(12) into an exactly known piece ${\cal{G}}^{(n)}({\cal{R}})$, containing the known RS-invariant
coefficients ${g}_{i}$, $(i=1,2,\ldots{n-1})$, and the unknown remainder
${\bar{\cal{G}}}^{(n)}({\cal{R}})$,
\ba
{\cal{G}}({\cal{R}})&=&{\cal{G}}^{(n)}({\cal{R}})+{\overline{\cal{G}}}^{(n)}({\cal{R}})
\nonumber \\
&=&(1+\sum_{k=1}^{n-1}{g}_{k}{\cal{R}}^{k})+\sum_{k=n}^{\infty}{\bar{g}}^{(n)}_{k}{\cal{R}}^{k}\;.
\ea
Standard ${\rm{N}}^{n}$LO ${\alpha}_{s}(\mu)$BPT is exactly
equivalent to ${\Lambda}$BPT with a particular RS-dependent choice for the
unknown ${\bar{g}}^{(n)}_{k}$ coefficients in Eq.(36). These follow
on replacing the RS-invariants ${\rho}_{n+1},{\rho}_{n+2},\ldots$, by RS-dependent
${\bar{\rho}}^{(n)}_{n+1}, {\bar{\rho}}_{n+2}^{(n)},\ldots$ in which ${r}_{n+1},{r}_{n+2},\ldots$,
have been set to zero in the expressions for the ${\rho}_i$. Thus at NLO level one has
from Eqs.(8)
\ba
{\bar{\rho}}_{2}^{(1)}&=&-{r}_{1}c-{r}_{1}^{2}
\nonumber \\
{\bar{\rho}}_{3}^{(1)}&=&{r}_{1}^{2}c+4{r}_{1}^{3}\;.
\ea
Correspondingly, using Eqs.(13) one finds expressions for the ${\bar{g}}^{n}_{i}$,
\ba
{\bar{g}}_{1}^{(1)}&=&-\frac{({r}_{1}^{2}+{r}_{1}c)}{b}
\nonumber \\
{\bar{g}}_{2}^{(1)}&=&-\frac{(4{r}_{1}^{3}-{r}_{1}{c}^{2})}{b}+
\frac{({r}_{1}^{4}+2{r}_{1}^{3}c+{r}_{1}^{2}{c}^{2})}{{b}^{2}}\;.
\ea
At NNLO we would have
\be
{\bar{\rho}}_{3}^{(2)}=2{r}_{1}^{3}-4{r}_{1}{r}_{2}-{r}_{1}^{2}c-2{r}_{1}{\rho}_{2}\;.
\ee

As a concrete illustration let us suppose that ${r}_{1}=10$, a value which is
typical of the NLO corrections at the physical scale ${\mu}=Q$ for ${e}^{+}{e}^{-}$
jet observables such as thrust. Taking ${N_f}=5$ we then find that standard NLO
${\alpha}_{s}(\mu)$BPT at the physical scale is precisely
equivalent to ${\Lambda}$BPT with
\be
{\Lambda}_{\overline{MS}}=Q(0.06415){\cal{F}}({\cal{R}})[1+29.374{\cal{R}}-607.92{\cal{R}}^{2}+\ldots]\;.
\ee
The terms in the square bracket correspond to $1+{\bar{\cal{G}}}^{(1)}({\cal{R}})$. It is seen that the
coefficients ${\bar{g}}^{(1)}_{i}$ are large. In fact at NLO it is straightforward to
write down the all-orders result for ${\bar{\cal{G}}}^{(1)}({\cal{R}})$,
in closed form, one finds \cite{r9}
\be
{\Lambda}_{\overline{MS}}=Q{\cal{K}}_{\cal{R}}^{\overline{MS}}{\cal{F}}({\cal{R}}){\exp}\left[F({\cal{R}})-
F\left(\frac{-1+\sqrt{1+4{r}_{1}{\cal{R}}}}{2{r}_{1}}\right)+{r}_{1}\right]\;.
\ee
If we take ${\cal{R}}=0.05$, which is typical for ${e}^{+}{e}^{-}$ event shape
observables at $Q={M}_{Z}$, then we find that $1+{\bar{\cal{G}}}^{(1)}(0.05)=14.97$,
to be compared with the value ${\cal{G}}({\cal{R}})=1$ which is all that is exactly known
from a NLO perturbative calculation. The apparently large higher-order contributions
to ${\cal{G}}({\cal{R}})$ result from a particular choice of RS, and there is no
reason to believe that the unknown coefficients ${\bar{g}}_{i}$ really
are large. These coefficients are simply unknown and are completely independent
of the RS. One therefore runs the risk with ${\alpha}_{s}(\mu)$BPT that misleading
estimates of $\Lambda$ are obtained. We believe that a more sensible approach given
an exact ${\rm{N}}^{n}$LO calculation, is
to approximate ${\cal{G}}({\cal{R}})$ by ${\cal{G}}^{(n)}({\cal{R}})$. If in fact
the uncalculated remainder ${\bar{{\cal{G}}}}^{(n)}({\cal{R}})$ is large then
${\rm{N}}^{n}$LO perturbation theory is not adequate to compute the observable.
If it is reasonably small then ${\Lambda}$BPT will give a good estimate of
the actual ${\Lambda}_{\overline{MS}}$ value. In contrast using the physical scale
if $r$ turns out to be sizeable is equivalent to assuming values for the unknown
${g}_{i}$ higher order coefficients which are so large that fixed-order
perturbation theory would be invalid, for instance with ${\cal{R}}=0.05$ the first
few terms in the expansion of ${\cal{G}}({\cal{R}})$ in the square bracket in Eq.(40)
are $[1+1.469-1.520+\ldots]$. By studying the scatter in the ${\Lambda}_{\overline{MS}}$
extracted with NLO $\Lambda$BPT from QCD observables one is directly learning about the
relative size of the unknown NNLO RS invariant ${\rho}_{2}$. Applying this method
to a selection of ${e}^{+}{e}^{-}$ event shape observables in Ref.\cite{r9} led to
the estimate $|{\rho}_{2}|\approx{50}$. As pointed out in Ref.\cite{r15} further
information about higher-order corrections can also be gleaned from studies of
the running of observables with energy $Q$. For instance if ${\rho}_{2}$ is small
then the running of the observable should be well-approximated by the NLO result
\be
Q\frac{d{\cal{R}}}{d{\ln}Q}=-b{{\cal{R}}}^{2}(1+c{\cal{R}})\;.
\ee

We cannot resist one further comment. If an observable has been computed at NLO
level so that $r_1$ is known, a part of the next coefficient ${r}_{2}$ can be
predicted. Rearranging Eq.(8) one finds
\be
{r}_{2}=({r}_{1}^{2}-{r}_{1}c+{c}_{2})-{\rho}_{2}\;.
\ee
The bracketed combination is an ``RG-predictable'' part of the NNLO coefficient
which is known given a NLO calculation. If the ``RG-unpredictable'' piece given
by the NNLO RS invariant ${\rho}_{2}$ is small and ${r}_{1}$ is reasonably large
then ${r}_{2}$ may be well-estimated by its RG-predictable part. This method has
been widely applied to make estimates of higher-order coefficients in QCD
perturbation theory \cite{r16}, and these estimates have often proved reasonably
accurate. We would argue that such estimates are pointless in the sense that
if ${r}_{2}$ can be accurately estimated in this way then ${\rho}_{2}$ is small
and NLO $\Lambda$BPT will potentially give a good estimate of the actual ${\Lambda}_{\overline{MS}}$.
${\Lambda}$BPT in fact resums RG-predictable terms to {\it all-orders}. For instance if
we make the simplification that $c=0$ the RG-predictable estimate for ${r}_{n}$ is
simply ${r}_{1}^{n}$. Summing this geometric progression to all-orders assuming the
one-loop form for the coupling $a(Q)=1/b{\ln}(Q/{\tilde{\Lambda}}_{\overline{MS}})$ leads \cite{r9,r11}
to the NLO $\Lambda$BPT result
\be
{\cal{R}}(Q)=\frac{1}{b{\ln}(Q/{\Lambda}_{\cal{R}})}\;.
\ee
This all-orders resummation of RG-predictable terms is precisely equivalent
to the complete resummation of ultraviolet logarithms advocated in the CORGI
approach \cite{r11}.

The first two sections have reviewed and amplified previous work \cite{r9,r11}. In
the remaining sections we turn to some new extensions of $\Lambda$BPT.
\section*{4 $\Lambda$BPT with massive quarks}
If quarks with pole masses $\{{m}_{j}\}$ are present then the dimensional analysis statement of
Eq.(6) becomes
\be
\frac{\partial{\cal{R}}}{\partial{{\ln}Q}}=f({\cal{R}}(Q),\{{m}_{j}\}/Q)\;.
\ee
This is now no longer a separable differential equation, and so it is unclear how
to integrate it and recover the $\Lambda$BPT relation between ${\cal{R}}$ and
${\Lambda}_{\overline{MS}}$. In Ref.\cite{r17} the generalisation of Eq.(6) to
the case of massive quarks has been considered. One now has
\be
\frac{\partial{\cal{R}}}{\partial{{\ln}Q}}=-{f}_{0}{\cal{R}}^{2}(1+\sum_{n\geq{1}}{f}_{n}{\cal{R}}^{n})\;,
\ee
where the ${f}_{n}(\{{m}_{j}\}/Q)$ are RS-invariant coefficients which are $Q$-dependent
unlike the ${\rho}_{i}$ invariants in Eq.(7) encountered in the massless case.
To derive expressions for them one can consider the RG equation \cite{r17}
\be
\frac{\partial{\cal{R}}}{\partial{\ln}Q}+\sum_{i}\frac{\partial{\cal{R}}}{\partial{\ln}{m}_{i}}=
\beta(a)\frac{\partial{\cal{R}}}{\partial{a}}\;.
\ee
Here the sum is over quark flavours $i=1,2,\ldots,{N}_{f}$.
The series expansion of the ${\beta}(a){\partial{\cal{R}}}/{\partial{a}}$ term in
powers of ${\cal{R}}$ must coincide with that in Eq.(7) in the massless quark
limit when the second term vanishes, in which case ${f}_{i}\rightarrow{\rho}_{i}$,
\be
{\beta}(a)\frac{\partial{\cal{R}}}{\partial{a}}=-b{\cal{R}}^{2}(1+c{\cal{R}}+\sum_{n>1}{\rho}_{n}{\cal{R}}^{n})\;.
\ee
With massive quarks the ${\rho}_{n}$ have a dependence on $\{{m}_{j}\}/Q$ and are in general
RS-dependent. The mass derivative in Eq.(47) will have a series expansion
\be
\frac{\partial{\cal{R}}}{\partial{\ln}{m}_{i}}={h}_{0i}{\cal{R}}^{2}(1+\sum_{n\geq{1}}{h}_{ni}{\cal{R}}^{n})\;.
\ee
The ${h}_{ni}$ are again functions of $\{{m}_{j}\}/Q$ and in general depend on the RS. They
can be easily calculated by simply substituting the perturbation series in powers of $a$
for ${\cal{R}}$ on both sides of Eq.(49), and equating corresponding powers of $a$. One
finds \cite{r17}
\ba
{h}_{0i}&=&\frac{\partial{r}_{1}}{\partial{\ln}{m}_{i}}
\nonumber \\
{h}_{0i}{h}_{1i}&=&\frac{\partial}{\partial{\ln}{m}_{i}}({r}_{2}-{r}_{1}^{2})
\\
\vdots
\nonumber
\ea
Substituting these series expansions into Eq.(47), and equating powers of ${\cal{R}}$
one easily obtains expressions for the ${f}_{n}$ RS invariants in terms of the
${\rho}_{n}$ and ${h}_{0i}{h}_{ni}$ \cite{r17},
\ba
{f}_{0}&=&b+\sum_{i}{h}_{0i}
\nonumber \\
{f}_{0}{f}_{n}&=&b{\rho}_{n}+\sum_{i}{h}_{0i}{h}_{ni}\;,
\ea
where for $n=1$ ${\rho}_{1}=c$.\\

For general renormalisation
schemes the ${h}_{0i}{h}_{ni}$ and ${\rho}_{n}$ are separately RS-dependent, with
the RS-dependence cancelling between the two terms in Eqs.(51) to yield
RS-invariant expressions for the ${f}_{n}$. However, as we shall now show, for
mass-independent renormalisation schemes such as ${\overline{MS}}$ where the
beta-function coefficients $b,c,{c}_{2},\ldots$ in Eq.(4) are independent of
the quark masses $m_i$, the ${h}_{0i}{h}_{ni}$ and ${\rho}_{n}$ are {\it both}
RS-invariant. To see this consider the equation for the RS-invariant ${f}_{1}$.
Since $b$ is a universal RS-invariant in mass-independent schemes it immediately
follows that $\sum_{i}{h}_{0i}$ is RS-invariant. It is then clear that for each $i$ {\it individually}
${h}_{0i}$ must be RS-invariant since any ${m}_{i}/{\mu}$ dependence must cancel for each
term in the sum {\it separately}. Considering the expression for the RS-invariant
${f}_{1}$ and noting that ${\rho}_{1}=c$ is also a universal RS-invariant in mass-independent
schemes it follows that ${h}_{0i}{h}_{1i}$ is RS-invariant. The crucial point is now
that ${\rho}_{2}$ can be related to these ${h}_{0i}$ and ${h}_{0i}{h}_{1i}$ RS-invariants,
one has
\be
\frac{\partial{\rho}_{2}}{\partial{\ln}{m}_{i}}={h}_{0i}{h}_{1i}-c{h}_{0i}\;.
\ee
The fact that the mass derivative of ${\rho}_{2}$ is RS-invariant is sufficient to
demonstrate that ${\rho}_{2}$ is RS-invariant. Any RS-dependence in ${\rho}_{2}$ would
have to be independent of the ${m}_{i}$, and would hence persist even with
massless quarks, but this is excluded because ${\rho}_{2}$ is an RS-invariant
in the massless case. Considering the expression for the RS-invariant ${f}_{2}$
the fact that ${\rho}_{2}$ is RS-invariant then implies that ${h}_{0i}{h}_{2i}$
is RS-invariant. One can then express the mass derivative of ${\rho}_{3}$ in terms of
${h}_{0i},{h}_{0i}{h}_{1i},{h}_{0i}{h}_{2i}$ showing that ${\rho}_{3}$ is RS-invariant,
and so continuing in this iterative fashion one establishes that the ${h}_{0i}{h}_{ni}$
and the ${\rho}_{n}$ are all RS-invariants in mass-independent RS's. The fact that
the ${\rho}_{n}$ are RS-invariant in mass-independent schemes makes it
plausible that the ${\Lambda}$BPT arising from integrating Eq.(45) has the same
form as in the massless case, with the ${\cal{K}}_{\cal{R}}$ and ${g}_{i}$
involving the same combinations of perturbation theory and beta-function coefficients.
This is indeed the case as we shall now show.\\

The differential equation in Eq.(45) is no longer separable with quark masses and we cannot directly
integrate it to obtain the ${\Lambda}$BPT relation as in the massless case. We proceed
by noting that from Eq.(9) in the massless case we have (assuming ${\overline{MS}}$ subtraction)
\be
{\ln}\frac{Q}{{\Lambda}_{\overline{MS}}}=K{\ln}{\cal{R}}+\frac{{K}_{-1}}{{\cal{R}}}+\sum_{n\geq{0}}{K}_{n}{\cal{R}}^{n}\;,
\ee
where the ${K}_n$ are RS-invariant and $Q$-independent coefficients, the first few being
\be
K=\frac{c}{b}\;\;,{K}_{-1}=\frac{1}{b}\;\;,{K}_{0}=\frac{r}{b}+c{\ln}c\;\;,{K}_{1}=\frac{{\rho}_{2}}{b}-\frac{c^2}{b}\;,\ldots
\ee
With massive quarks one expects a similar expression but with the $K_i$ now having a
dependence on $\{{m}_{j}\}/Q$. To evaluate the ${K}_{i}$ with massive quarks one differentiates
both sides of Eq.(53) $\frac{\partial}{\partial{\ln}Q}$ to obtain
\ba
1&=&-{f}_{0}{\cal{R}}^{2}(1+{f}_{1}{\cal{R}}
+{f}_{2}{\cal{R}}^{2}+\ldots)(\frac{K}{{\cal{R}}}-\frac{{K}_{-1}}{{\cal{R}}^{2}}
+{K}_{1}+2{K}_{2}{\cal{R}}+\ldots)
\nonumber \\
&+&({K}^{\prime}{\ln}{\cal{R}}+\frac{{K}_{-1}^{\prime}}{{\cal{R}}}+{K}_{0}^{\prime}+{K}_{1}^{\prime}{\cal{R}}+\ldots)
\;,
\ea
where we have used Eq.(46) and ${K}_{i}^{\prime}\equiv {\partial}{K}_{i}/{\partial}{\ln}Q$. Equating coefficients of
${\ln}{\cal{R}}$, ${\cal{R}}^{-1}$, ${\cal{R}}^{0}$, ${\cal{R}}\;\ldots$, on both sides of this equation leads
, respectively, to the following equations
\be
{K}^{\prime}=0\;,{K}_{-1}^{\prime}=0\;,{f}_{0}{K}_{-1}+{K}_{0}^{\prime}=0\;,
-{f}_{0}K+{f}_{0}{f}_{1}{K}_{-1}+{K}_{1}^{\prime}=0\;,\ldots\;.
\ee
In each case one has a first-order differential equation for ${K}_{i}$ which needs to be
integrated $\int{d}{\ln}Q$ to obtain ${K}_{i}$. The boundary condition is that in the massless
limit (${m}_{i}\rightarrow{0}$ or $Q\rightarrow{\infty}$) one must reproduce the coefficients in
Eq.(54). Integrating ${K}^{\prime}=0$ gives $K=A$, where $A$ is an arbitrary constant of integration, the
boundary condition then fixes $A=c/b$ and the massless result $K=c/b$ is reproduced. Similarly
${K}_{-1}^{\prime}=0$ integrates up to the massless result. Using Eq.(51) for $f_0$ and setting
${K}_{-1}=1/b$ the next equation becomes
\be
{K}_{0}^{\prime}=-\frac{1}{b}\sum_{i}\frac{\partial{r}_1}{\partial{\ln}{m}_{i}}\;.
\ee
Noting that ${h}_{0i}$ is RS-invariant and will be a function of the $\{{m}_{j}\}/Q$ we can apply the general result that
\be
\sum_{i}\frac{\partial{F}(\{{m}_{j}\}/Q)}{\partial{\ln}{m}_{i}}=-\frac{\partial{F}(\{{m}_{j}\}/Q)}{\partial{\ln}Q}
\ee
to recast Eq.(57) in the form
\be
{K}_{0}^{\prime}=\frac{1}{b}\frac{\partial{r}_{1}}{\partial{\ln}Q}\;.
\ee
This integrates trivially to give ${K}_{0}=({r}_{1}/b)+A$, where again $A$ is an arbitrary integration
constant. If we use the notation ${r}_{1}^{(0)}$ to denote ${r}_{1}$ in the limit of massless
quarks then we have
\be
{r}_{1}={r}_{1}^{(0)}+{\hat{r}}_{1}(\{{m}_{j}\}/{Q})\;,
\ee
where ${\hat{r}}_{1}$ contains the $\{{m}_{j}\}/Q$ dependence and vanishes in the massless limit.
As we discussed above in a mass-independent scheme ${h}_{0i}$, the logarithmic mass derivative
of ${r}_{1}$, is RS-invariant, i.e. $\mu$-independent. This implies that ${\hat{r}}_{1}$ is also
RS-invariant. Thus as in the massless case changes in scale correspond to translating
${r}_{1}$ by a constant, and the integration constant $A$ can be fixed so that
${K}_{0}=(r/b)+c{\ln}c$ reproducing the massless result. We now turn to the equation for ${K}_{1}^{\prime}$.
Substituting in the results for ${f}_{0}$ and ${f}_{0}{f}_{1}$ from Eqs.(51), and setting
$K=c/b$ and ${K}_{-1}=1/b$, use of the result in Eq.(58) to change from $\sum_{i}(\partial/\partial{m}_{i})$
to $-\partial/\partial{\ln}Q$ yields
\be
{K}_{1}^{\prime}=\frac{1}{b}\frac{\partial}{\partial{\ln}Q}[{r}_{2}-{r}_{1}^{2}-{r}_{1}c]\;.
\ee
This integrates up to give ${K}_{1}=[({r}_{2}-{r}_{1}^{2}-{r}_{1}c)/b]+A$, where $A$ is the
constant of integration. The boundary condition then fixes $A={c}_{2}-{c}^{2}/b$
to reproduce the massless result ${K}_{1}=({\rho}_{2}-{c}^{2})/b$.\\

A succinct proof that to all-orders the massless expressions for $K_i$ are reproduced with
massive quarks is as follows. Instead of differentiating both sides of Eq.(53) $\frac{\partial}{\partial{\ln}Q}$
one can instead differentiate $\sum_{i}\frac{\partial}{\partial{\ln}{m}_{i}}$. Equating coefficients
on both sides produces a set of conditions equivalent to Eqs.(56) with the replacements
${f}_{0}{f}_{n}\rightarrow{-}\sum_{i}{h}_{0i}{h}_{ni}$, and where ${K}_{n}^{\prime}$ is
replaced by $\sum_{i}\frac{\partial{K}_{n}}{\partial{\ln}{m}_{i}}$. But using the result
of Eq.(58) one sees that these new conditions are precisely equivalent to the original conditions of Eqs.(56)
with all terms involving the $b{\rho}_{n}$ part of ${f}_{0}{f}_{n}$ removed. This in turn means
that this ${b}{\rho}_{n}$ subset of terms must separately vanish, but this condition is just
the equation that determines a given ${K}_{i}$ in terms of the ${\rho}_{n}$ in the case of
massless quarks. So, as an example, we can determine $K_1$ by integrating Eq.(61) as we have seen.
We can also consider the next condition in Eqs.(56) involving ${K}_{2}^{\prime}$,
\be
-{f}_{0}{f}_{1}K+{f}_{0}{f}_{2}{K}_{-1}-{f}_{0}{K}_{1}+{K}_{2}^{\prime}=0\;.
\ee
Splitting this equation into the terms involving $b{\rho}_{k}$ and ${h}_{0i}{h}_{ki}$
 and demanding, as argued above, that each part separately vanishes, gives the
two conditions
\ba
-{c}^{2}+{\rho}_{2}-b{K}_{1}&=&0
\nonumber \\
\sum_{i}(-\frac{c}{b}{h}_{0i}{h}_{1i}+\frac{1}{b}{h}_{0i}{h}_{2i}-{h}_{0i}{K}_{1})+{K}_{2}^{\prime}&=&0\;.
\ea
The first condition determines ${K}_{1}$ to have the same form as the massless quark result.
In general the $b{\rho}_{i}$ subset of
terms in the differential equation for ${K}_{n}^{\prime}$ build
the condition satisfied by ${K}_{n-1}$ in the massless quark case.
Notice that for mass-dependent schemes such as momentum subtraction
the above proof fails because the result of Eq.(58) no longer holds since
now $F(\{{m}_{j}\}/{\mu},Q/{\mu})$, where $\mu$ is the
renormalisation scale. Crucially in the mass-independent case the
RS-invariance of the ${h}_{0i}{h}_{ni}$ underwrites the validity
of Eq.(58).\\

We turn in the next section to an attempt to resum large infra-red
logarithms for ${e}^{+}{e}^{-}$ event shape observables in the
${\Lambda}$BPT approach, and a phenomenological analysis to extract
the apparent size of power corrections for the measured thrust
and heavy jet mass distributions at various energies.

\section*{5 Event Shape Observables in $\Lambda$BPT}

Event shape variables provide some of the most interesting and useful ways to confront QCD calculations
with experiment (for a recent review see \cite{Dasgupta:2003iq}). Their infrared and collinear safety
guarantees that they can be calculated in QCD perturbation theory, but fixed order calculations describe
their distributions rather poorly.  This situation can be improved by the recognition that for a shape variable $y$,
vanishing in the 2-jet limit, large logarithms $L=\log(1/y)$
appear at each order of perturbation theory and must be resummed.  However, a full understanding of the
observed distributions requires the introduction of large {\it non-perturbative} effects, {\it power corrections}
$\propto e^{-b/a} \simeq \Lambda/Q$ where $Q$ is some hard scale (e.g. the  $e^+ e^-$ centre of mass energy).
Although the existence of such effects can be motivated by considering simple models
of hadronization \cite{Webber:1994zd} or through a renormalon analysis \cite{Nason:1995hd}, their magnitudes are not at present calculable in a truly
systematic way from the QCD Lagrangian. Therefore to fit the data we require the introduction of either a phenomenological hadronization
model or additional non-perturbative parameters.  Although this hampers attempts
to extract reliable measurements of $\alpha_S$ from the data, it also provides a good opportunity to study the IR behaviour
of QCD experimentally.  For example, in \cite{Dokshitzer:1995zt} it was proposed to relate the magnitude of the
$1/Q$ power correction to event shape means to the average value of a hypothetical infrared-finite coupling.  This approach
leads to predictions for $1/Q$ power corrections to all event shape means in terms of a {\it single} additional parameter,
$\overline{\alpha}_0(\mu_I)$, the zeroth moment (i.e. mean) of the coupling at scales  $0<\mu<\mu_I$ (typically
$\mu_I \simeq 2$GeV).  In \cite{Dokshitzer:1997ew} it was shown how this approach could be extended to apply to event shape distrubutions.
Since then many experimental studies have appeared, fitting event shape means and distributions simultaneuosly for
$\alpha_{\MSb}(M_Z)$ and $\alphazero(\mu_I)$.  Generally an approximate (up to corrections $\simeq 25\%$) universality of the $\alphazero$ values
is observed, supporting the hypothesis that power corrections can be related to a universal coupling in this way.  However,
all these fits use the $\MSb $ scheme with the ``physical'' scale choice $\mu=Q$.  For the event shape means an obvious
alternative is to work within the $\Lambda$BPT framework described here.  This was first carried out for 1-thrust in
\cite{Campbell:1998qw}, and somewhat reduced power corrections were found compared to the physical scale approach.  This suggests the possibility that the
power suppressed effects are partly compensating for missing higher order perturbative terms.      Recently a more extensive
analysis was performed by the DELPHI collaboration \cite{r12}, taking into account effects arising from the finite bottom quark mass
via Monte Carlo simulations.  They found remarkably small power corrections within the $\Lambda$BPT approach, which for many observables were
consistent with zero.  Indeed, the $\Lambda$BPT predictions with {\it no} power corrections whatsoever gave a better description of the data
than the model of \cite{Dokshitzer:1995zt,Dokshitzer:1997ew} with a universal $\alphazero$.  In light of these surprising results, it is interesting to
consider applying the $\Lambda$BPT method to the event shape distributions.

In fact, event shape distributions have previously been studied within the $\Lambda$BPT framework \cite{r9}. To do this
an effective charge was constructed separately for each bin of the data, and
NLO QCD calculations were used to extract $\Lambda_{\MSb}$ at centre of mass energy $Q=M_Z$.  Non-perturbative
effects were taken into account by using Monte Carlo simulations to correct the data back to ``parton-level''
distributions. This generally improved the quality of the prediction, but with this approach it is hard to see
whether the $\Lambda$BPT distributions prefer smaller hadronization corrections than the $\alpha(\mu)$BPT ones.
Moreover, even after these corrections were applied there were still two kinematical
regions where the effective charge ceased to be a good description of the data: the
2-jet limit where large logs enhance the higher-order perturbative coefficients, and the region (which exists
for many observables) where the LO result vanishes, causing $r_1 \rightarrow \infty$.  The latter problem
is hard to address within the effective charge approach, but the former problem can in principle be
alleviated by introducing a resummation of the effective charge beta function.  The basic idea is that large
logs appearing in the perturbation series for the distribution lead to large logs in the perturbation
series for $\rho(\mathcal{R})$ \cite{r15}.  We can then resum this series to arrive at an improved description of the data in the 2-jet region
of phase space.

In this section we first show the effect of replacing the {\it hadronization corrections} of \cite{r9}
with an analytical power correction ansatz.  For simplicity, we use a shift in the distribution by an amount $C_1/Q$.
This form can be motivated by considering simple models of hadronization or through a renormalon analysis \cite{Webber:1994zd} and
has been found successful phenomenologically (see for example \cite{r12}).  Although better fits are often obtained using the model
of \cite{Dokshitzer:1995zt,Dokshitzer:1997ew}, because we are using a different perturbative approximation to standard NLO QCD, the
subtractions needed to remove double counting will not in general be so simple.
Next, we outline how we can arrive at an arbitrarily accurate numerical approximation to the resummed
effective charge beta function $\rho(\mathcal{R})_{\rm NLL}$ by manipulation of the
standard NLL resummed expressions.  We then present results showing the effect of using this modified
$\rho$ on fits to the distributions of 1-thrust ($\tau\equiv 1-T$) and heavy jet mass ($\rho_h$).
We use data taken over a wide range of centre-of-mass energies $Q=35-189$GeV (Refs.[26-37]). Lacking information on the correlation
between data points we have simply combined statistical and systematic errors in quadrature and performed a min-$\chi^2$ fit, allowing
$\chi^2$ to vary by 4 from its minimum to estimate a $2\sigma$ error.  This over-simplistic treatment means that our errors cannot
be considered reliable, however the central values of $\Lambda_{\overline{MS}}$ do give an impression of the effect of including the
power corrections and logarithmic resummation into the $\Lambda$BPT framework.

Before proceeding it is worth commenting on an objection often raised to the appliction of the ``method of effective charges'' to
exclusive quantities such as event shape distributions.  The idea is that the dependence of the physical quantity
on multiple scales invalidates the dimensional analysis argument as presented here in Section 2.  However, as pointed out in \cite{r10}, this is not
really the case.  Given an observable $\mathcal{R}=\mathcal{R}(Q_1,Q_2,...,Q_n)$ depending on $n$ scales, one can simply re-express it as
$\mathcal{R}=\mathcal{R}(Q_1,Q_2/Q_1,...,Q_n/Q_1) \equiv \mathcal{R}_{x_2,...,x_n}(Q_1)$.  Here the $x_i \equiv Q_i/Q_1$
are {\it dimensionless} quantities that can be thought
of as labelling the effective charge which is now a function of one single dimensionful scale $Q_1$.  We can then
write an effective charge beta function for this $\mathcal{R}$ describing the energy evolution of our physical observable for
fixed values of the ratios $x_i$.  Of course, this formal manipulation cannot tell us whether the $\rho$ function we
arrive at in this way will be well approximated by its NLO terms, which is what we require for most phenomenological
applications, given the current state of the art in perturbative QCD calculations.  One reason in particular why this
might not be the case is if some of the $x_i$ become large - typically this
leads to powers of large logs $L_i=\log(x_i)$ enhancing the coefficients $r_n$ in the perturbative series for $\mathcal{R}$.  These
logs can then appear also in the $\rho_n$ and invalidate the simple NLO form for $\rho$.  To solve this we need to apply
some form of resummation, which is precisely our goal in this analysis.  However, as with any perturbative analysis (with an
unknown remainder function), we can only really determine the accuracy of our approximations {\it a posteriori} by examining
the quality and consistency of fits to data.

Let us now consider the effects of analytical power corrections on the results of \cite{r9}.
The procedure used in \cite{r9} was to write an effective charge to represent the value of the event shape distribution
integrated over each bin of the data.  First, the Monte Carlo program EERAD \cite{Giele:1991vf} was used to compute the NLO perturbative coefficients
\footnote{ For our analysis we actually used EVENT2 \cite{Catani:1996jh} and we have checked that both programs give consistent coefficients.}
for each bin
\begin{equation}
\label{eq:NLOevtshape}
\int_{bin~i} dy ~ \frac{1}{\sigma} \frac{d\sigma}{dy}  = A_i \alpha + B_i \alpha^2 + O(\alpha^3).
\end{equation}
These were then used to write an effective charge, from which a value for $\Lambda_{\MSb}$ could be extracted by feeding the data into
(\ref{eq:LambdaExtractor}).  Here, to introduce a fit for $C_1$ we drop this ``direct extraction'' approach and instead
perform a minimum $\chi^2$ fit for $\Lambda_{\MSb}$ and $C_1$.  For this to work, we need to exclude the regions where the EC approach
cannot fit the data.  For comparison with \cite{r9}, we choose the same ranges selected there (based on
the flatness of $r_1$), except that the lower end of the range is made proportional to $1/Q$ when looking at data away from $Q=M_Z$.
The reason for this is that sub-leading non-perturbative effects are expected to become important for $y \simeq \Lambda/Q$ \cite{Dokshitzer:1997ew}.
As we are shifting the predictions before comparing to data we require the NLO coefficients evaluated for arbitrary bin edges.  We have
approximated these using a set of order 6 polynomial interpolations from the output of EVENT2.  We have checked by halving the Monte Carlo bin size to 0.005
that this induces no sizeable error (using the doubled sampling changes the best fit values here by less than 2\%).

\begin{figure}
\begin{center}
\includegraphics[scale=0.5]{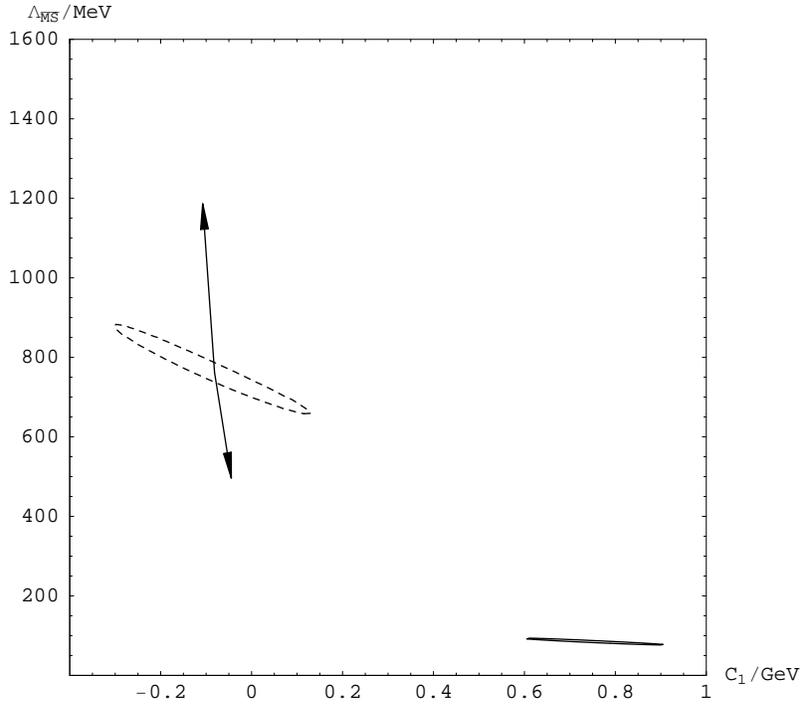}
\end{center}
\caption{1-thrust: Fits for $\Lambda_{\MSb}$ and $C_1$
within the framework of \cite{r9} (solid ellipse) and standard NLO QCD peturbation theory (dashed ellipse).
In the latter case the scale is chosen to be $\mu=Q$, and the effect on the central value of a change of renormalization scale by a factor of 2 is indicated by the arrows.
2$\sigma$ error contours are shown (from allowing $\chi^2$ to vary within 4 of its minimum).  The fit range is $1-T=0.055M_Z/Q-0.23$.}
\label{fg:burbyThrust}
\end{figure}

\begin{figure}
\begin{center}
\includegraphics[scale=0.5]{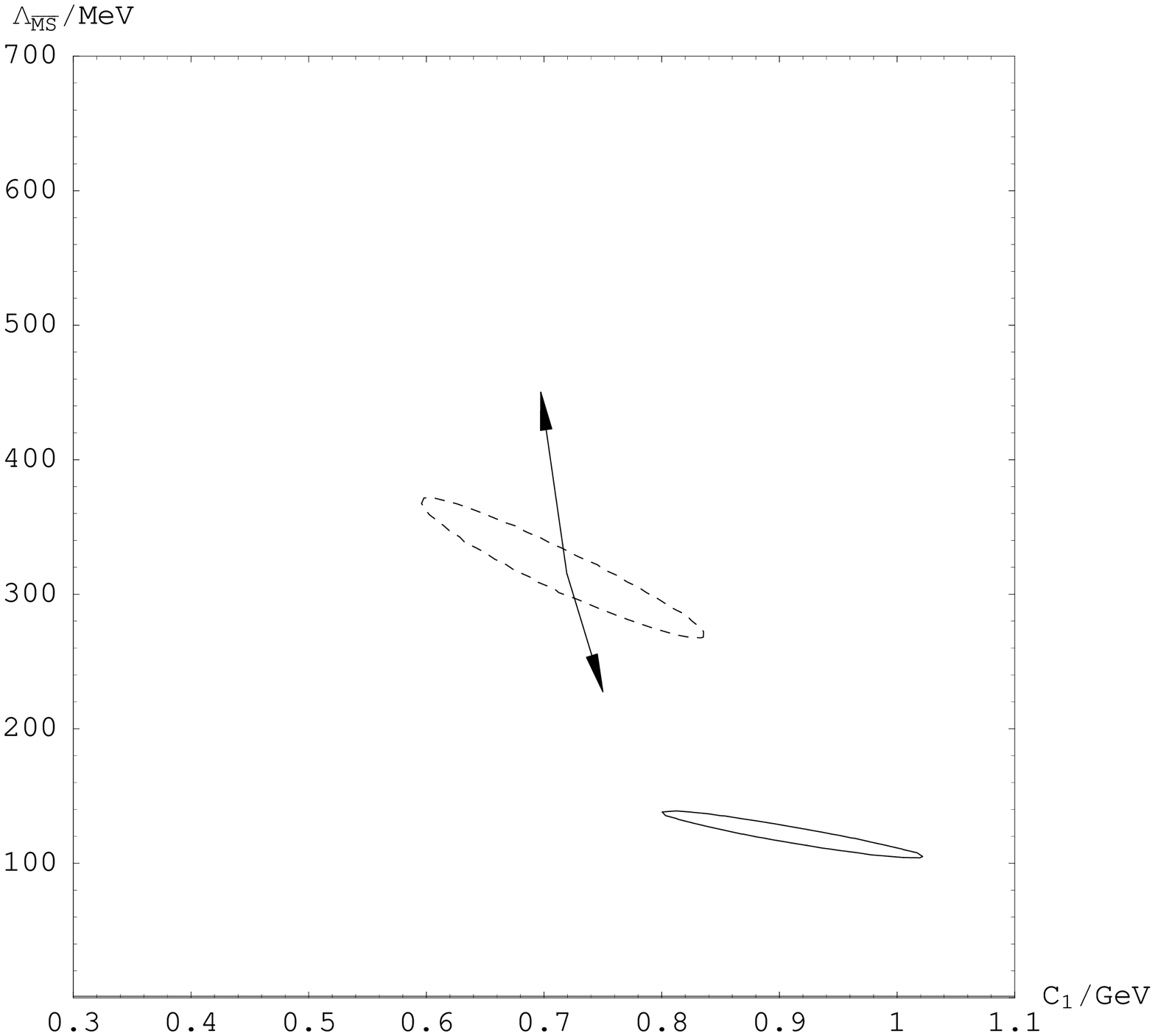}
\end{center}
\caption{As \reffg{burbyThrust} but for heavy jet mass.  The fit range is $\rho_h=0.035M_Z/Q-0.2$. }
\label{fg:burbyHjm}
\end{figure}

The results for 1-thrust and heavy jet mass are presented in \reffg{burbyThrust} and
\reffg{burbyHjm} respectively.

In the case of thrust it appears that the $\Lambda$BPT results prefer larger power corrections (and significantly smaller $\LMSb$ values).
For heavy jet mass, the situation is similar, although the differences are not quite as extreme.  However, in
both cases we find comparable $\Lambda_{\MSb}$ values to those found in \cite{r9} using hadronization corrected data rather than
an analytical power correction ansatz.
\footnote{
  However, the values of $\Lambda_{\MSb}$ quoted in \cite{r9} are actually wrongly normalized for two reasons.  Firstly the factor of
  $(2c/b)^{(c/b)} \simeq 0.85$ was omitted, so the results are really values for $\tilde{\Lambda}_{\MSb}$.  Secondly, the results
  of EERAD were normalized to the {\it Born} cross-section $\sigma_0$, whereas the data are normalized to the {\it total} cross-section
  $\sigma$, and this was not taken into account.  Mutiplying the EERAD perturbation series by a correction
  factor $\sigma_0/\sigma=1-\alpha/\pi+\cdots$ decreases $r_1$ by exactly 1,  increasing the extracted $\Lambda_{\MSb}$ values
  by $e^{1/b}$.  So the total correction factor to apply to the results of \cite{r9} is $(2c/b)^{(c/b)}e^{1/b} \simeq 1.11$.
}

Now, in preparation for applying our log resummations, we turn to the issue of handling exponentiation in the $\Lambda$BPT framework.
The typical form of an event shape distribution as a double expansion in $a$ and $L=\log(1/y)$ is
\begin{equation}
\frac{1}{\sigma} \frac{d\sigma}{dy} = A_{LL}(aL^2) + L^{-1} A_{NLL}(aL^2) + L^{-2} A_{NNLL}(aL^2) + \cdots\;,
\end{equation}
where the $A$ functions have a perturbative expansion $A(x) = {A}_{0}x+{A}_{1}{x}^{2}+\ldots$,
and the $A$ are known up to
NNLL accuracy.  However, some event shapes (including 1-thrust and heavy jet mass) exponentiate, meaning we can also write
\begin{eqnarray}
R_y(y') \equiv \int_0^{y'} dy \frac{1}{\sigma} \frac{d\sigma}{dy} = C(a\pi) \exp(L g_1(a\pi L) + g_2(a\pi L) + a g_3(a\pi L) + \nonumber \\
a g_3(a\pi L) + \cdots) + D(a\pi ,y)\;,
\end{eqnarray}
where $g_1$ and $g_2$ are known \cite{Catani:1991kz}.  When working with this form of the distribution it is conventional to refer to
$g_1$ as containing the {\it leading logarithms} and $g_2$ as containing the {\it next-to-leading logarithms}.
$C = 1+O(a)$ is independent of $y$, and $D$ contains terms that vanish as $y \to 0$.  These can be calculated to NLO by comparison
to fixed order results.  However, there is no unique way of including this fixed order information into the $R_y(y')$ (this is the
so-called matching ambiguity).  For example, it is also legitimate to include the $C, D$ terms into the exponent (termed ``log R matching''),
as the difference is of order $a^3$.  Here we choose to define
\begin{equation}
R_y(y') = \exp(r_0 \cal{R})\;,
\end{equation}
so that all the physics is encoded into a single effective charge.  This is similar to log R matching in that if we re-expand $r_0 \cal{R}$ in terms
of $a$ and $L$ the $C$ and $D$ functions will clearly appear in the exponent.  However, in this approach there is no {\it separate} matching ambiguity because
once we have picked the effective charge the inclusion of $C$ and $D$ is automatically determined.
In particular the full exact NLO coefficient $r_1$ in a given RS is reproduced if
${\cal{R}}$ which solves Eq.(16) is expanded in the coupling $a$ for that scheme, thanks to
the RS invariant ${\Lambda}_{\cal{R}}$ which appears on the lefthand side of the equation.

To perform a resummation of logs in $\rho(\cal{R})$ we first note that the $r_n$ have the form
\begin{equation}
r_n = r_n^{\rm LL} L^n + r_n^{\rm NLL} L^{n-1} + \cdots\;,
\end{equation}
so the structure of the $\rho_n$ as illustrated in Eq. (\ref{eq:rho_n}) implies that
\begin{equation}
\rho_n = \rho_n^{\rm LL} L^n + \rho_n^{\rm NLL} L^{n-1} + \cdots\;.
\end{equation}
Because $L$ is a logarithm of a physical quantity, this expansion of the $\rho_n$ is scheme invariant.
We can thus define resummed, scheme invariant approximations to $\rho(\cal{R})$ as follows
\ba
\rho_{\rm LL}(\mathcal{R})& = &-b \mathcal{R}^2 (1 + c \mathcal{R} + \sum_{n=2}^{\infty} \rho_n^{\rm LL} L^n \mathcal{R}^n ) \\
\rho_{\rm NLL}(\mathcal{R})& = &-b \mathcal{R}^2 (1 + c \mathcal{R} + \sum_{n=2}^{\infty} (\rho_n^{\rm LL} L^n + \rho_n^{\rm NLL} L^{n-1}) \mathcal{R}^n )\;.
\ea
Crucially $\rho_n^{\rm LL}$ is only a function of the $r_m^{LL}$, and $\rho_n^{\rm NLL}$ depends only
on the $r_m^{\rm LL}$ and $r_m^{\rm NLL}$.  This means that we can construct $\rho_{\rm NLL}$ given knowledge of
$\cal{R}_{\rm NLL}$.  In fact, it is a simple matter to arrive at a numerical approximation to both $\rho_{\rm LL}$ and $\rho_{\rm NLL}$ starting
from $\cal{R}_{\rm NLL}$.  To do this, we first construct the $\bar{\rho}$ corresponding to $\cal{R}_{\rm NLL}$.  Because the
NLL parts of this function only depend on the NLL parts of $\cal{R}_{\rm NLL}$ which are exact by definition, $\bar{\rho}$
agrees with $\rho_{\rm NLL}$ to NLL accuracy.  Then, simply truncating $\bar{\rho}$ by numerically taking limits
($L \to \infty$ with $L\cal{R}$ fixed) allows us to extract the LL and NLL terms.  In some sense the $c \mathcal{R}$ term is NLL as it
is ${\rm O}(L^{-1})$ in this limit, but we include it in $\rho_{\rm LL}$ as it is obviously present in the full expression, and this avoids
having to modify (\ref{eq:blogQfromFG}).  As an alternative to this numerical procedure we can evaluate the
first few terms of $\rho_{\rm NLL}$ analytically by series reversion and composition (in practice we use this form for
$\mathcal{R} \leq 0.005$ to ensure exact cancellation of the singularities between the two terms in $\mathcal{G}$).
These converge rapidly to the numerical, resummed results.  All of our calculations were carried out using
the computer algebra system {\it Mathematica} \cite{Mathematica}, allowing the use of arbitrary precision arithmetic
in taking the $L \to \infty$ limit.

These $\rho_{\rm NLL}$ and $\rho_{\rm LL}$ functions can be used to make predictions for $R(y)$, by inserting them
in Eq.(16) and numerically solving
the transcendental equation. It would in principle have been possible to perform corresponding resummations
for the ${\cal{G}}({\cal{R}})$ function, and then use the ${\Lambda}$BPT Eq.(24), but construction of ${\rho}({\cal{R}})$
is more straightforward.   By taking the difference in $R(y)$ across the
bins in each experimental data set a comparison to data can be carried out, including a $1/Q$ power correction by using
$R_{PC}(y) = R_{PT}(y - C_1/Q)$.  Once again, we need to choose the fit ranges so as to exclude the regions
where this effective charge description fails.  We now discuss this choice.

After exponentiation, the problem as $r_0 \to 0$ remains, and in fact
for thrust worsens; unfortunately this means we need to restrict the fits to $1-T < 0.18$, $\rho_h < 0.24$  to obtain good fits
in the $\Lambda$BPT approach. This suggests that writing an effective charge outside of the exponent as in \cite{r9}
might be more sensible; this is certainly possible and leads to a $\rho$ function where the LL, NLL and NNLL terms are known.
However this creates problems with the resummation.  To see this, note that in this case $\rho_{\rm LL}$ consists of just the
double logarithmic terms containing powers of $L^2 R$ and a $c R^3$ term and, apart from this $c R^3$ term, is identical to the $\bar{\rho}$ induced by the one-loop beta
function $\beta(a)=-ba^2$ and the double logarithmic distribution
\be
\frac{1}{\sigma} \frac{d\sigma}{dy} = \frac{d}{dy} \exp (- k L^2 a ) = 2 k e^{-L} L (a \exp (-k L^2 a)) \equiv 2 k e^{-L} L (\bar{\mathcal{R}})\;,
\ee
where $k$ is a constant ($4/3$ for thrust and heavy jet mass).  This distribution has a peak as a function of $a$ at $a_{max}=1/kL^2$, and so its
inverse only exists for $\bar{\mathcal{R}} < \bar{\mathcal{R}}(a_{max}) = e^{-1}/kL^2$.  As a consequence, for $c=0$, $\rho_{LL}(\mathcal{R})$ vanishes
at this point (where a branch cut starts).  Adding the $cR^3$ term back in, and later adding the rest of the NLL terms, does not remove this branch cut.
As $\mathcal{R}$ is evolved from $Q=\infty$ it increases until it reaches this maximum value, and then its evolution becomes undefined.
This value turns out to be too small to fit the data.  One could possibly ``switch branches'' of $\rho$ at this point
and allow $\mathcal{R}$ to decrease again, although this would
of course still not provide a good fit to the data.  Note also that this zero of $\rho$ does {\it not } correspond to an ``infrared freezing'' type
behaviour because $\rho$ approaches the zero as a fractional power of $\mathcal{R}-\mathcal{R}_{max}$ - thus the singularity in (\ref{eq:blogQfromFG}) is integrable
and the zero is reached after a finite amount of evolution in $Q$.
Accordingly we are forced into using the exponentiated form despite the problems when $r_0 \to 0$.

A lower cut on the shape variables is also necessary.  Although we expect
the behaviour of our distributions to be improved in the 2-jet region thanks to the resummation, the onset of non-perturbative effects
more complicated than a simple $1/Q$ shift means that we still need to impose a lower cut.  In fact, even
at the purely perturbative level there are problems with our resummed results in this region.  These are essentially due to the growth of
$\mathcal{R}$ as $r_1$ grows large (i.e. as $\Lambda_{\mathcal{R}}$ approaches $Q$).  This leads into a breakdown of the series of
logarithmic corrections to $\rho$ (as e.g. the NNLL terms are ${\rm O}(\mathcal{R})$ with respect to the NLL ones).  In fact, $\mathcal{R}$
eventually grows large enough that we encounter a branch cut in $\rho$ which appears due to the branch cut in $g_1$.  Clearly this behaviour is
unphysical, and must be avoided in our fits to data.
As discussed above we take the lower cut to be proportional to $1/Q$ and obtain good fits for $\rho_h,1-T>0.05M_Z/Q$.
Any bins not lying within this range have been left out of the fit; a summary of the data we actually used is given in Tables 3 and 4. We have also removed the JADE data at $35$ and
$44$ GeV from the heavy jet mass fits, since its inclusion dramatically worsens the fit
quality for all the predictions.

Results of these fits for thrust are shown in \reffg{thrust}, and for heavy jet mass in \reffg{hjm}.  For comparison,
fits using $\alpha_s(\mu)$BPT to the same data are also shown. The fit range for these could in principle
be extended as they do not suffer
from the $r_0 \to 0$ problem that afflicts the $\Lambda$BPT), however, in order to facilitate a
direct comparison between the two approaches we have used the same fit range for both.
The most notable
feature of the results is the stability of the $\Lambda_{\overline{MS}}$ values found within the $\Lambda$BPT
framework as we move from NLO to LL and then to NLL accuracy, while the fit quality hardly changes.
It seems that the effect of the leading and next-to-leading logarithms can be mimicked by an increase in $C_1$.
These $\Lambda_{\overline{MS}}$ values are, however, still smaller than the world average.
Some examples of the actual distributions are shown in \reffg{nloFits} and \reffg{nllFits}.

To investigate the sensitivity of these results to our choice of fit range we have redone the fits for a ``low'' range and
a ``high'' range.  The low range was determined by decreasing the upper cut until half the bins were excluded, and the
high range was determined by increasing the lower cut similarly.  The effects of these changes on the central values
of $\Lambda_{\overline{MS}}$ and $C_1$ are shown in Table 1 (for 1-thrust) and Table 2 (for heavy jet mass).

Lastly, we have also considered the so-called ``modification of the logs'' that is often invoked in studies of event shape
variables.  This consists of modifying $L=\log(1/y) \to \log((2y_{max}-y)/y)$ to ensure that the resummed parts of the expression
vanish at the upper kinematic limit $y_{max}$ (which is 0.5 for both $T$ and $\rho_h$).   The change in central values is shown in Tables 1 and 2. One finds that the fitted values change
very little. This is to be expected since the restricted fit range automatically
ensures that the logarithm is essentially unchanged in that region.

\begin{table}
\begin{tabular}{|l|c|c|c|}
\hline
Prediction & $\Lambda$/MeV & $C_1$/GeV & $\chi^2/dof.$ \\
\hline
NLO $\Lambda$BPT & 98,116,118 & 1.02,0.85,0.89 & 24/46,59/88,27/44\\
NLO $\alpha_S(\mu)$BPT & 446,463,517 & 1.47,1.42,1.28 & 26/46,57/88,23/44\\
\hline
LL $\Lambda$BPT & 101,119,108 & 0.80,0.63,0.88 & 23/46,63/88,30/44\\
LL $\alpha_S(\mu)$BPT & 371,417,478 & 1.12,1.02,0.83 & 24/46,49/88,22/44\\
\hline
NLL $\Lambda$BPT & 103,107,93 & 0.50,0.52,0.93 & 23/46,84/88,34/44 \\
NLL $\Lambda$BPT (mod) & 104,110,95 & 0.50,0.47,0.90 & 23/46, 81/88, 34/44 \\
NLL $\alpha_S(\mu)$BPT & 200,233,268 & 0.94,0.79,0.60 & 23/46, 49/88, 23/44\\
\hline
\end{tabular}\\
Table 1.  Sensitivity of our fit values for $\Lambda_{\overline{MS}}$ and $C_1$ to the choice
of fit range for thrust.  The fit ranges are $0.05M_Z/Q-0.1$ (low), $0.05M_Z/Q-0.18$ (normal), $0.11M_Z/Q-0.18$ (high).
\end{table}

\begin{table}
\begin{tabular}{|l|c|c|c|}
\hline
Prediction & $\Lambda$/MeV & $C_1$/GeV & $\chi^2/dof.$\\
\hline
NLO $\Lambda$BPT & 120,114,142 & 1.19,1.22,0.94 & 19/42,50/84,29/41 \\
NLO $\alpha_S(\mu)$BPT & 236,115,221 & 1.65,1.84,1.27 & 19/42,68/84,37/41\\
\hline
LL $\Lambda$BPT &  124,123,128 & 1.06,1.07,1.01 & 21/42,51/84,29/41\\
LL $\alpha_S(\mu)$BPT & 185,132,146 & 1.39,1.57,1.33 & 19/42,63/84,36/41\\
\hline
NLL $\Lambda$BPT & 125,127,122 & 0.99,0.97,1.04 & 21/42,51/84,29/41\\
NLL $\Lambda$BPT (mod) & 126,128,122 & 0.98,0.97,1.05 & 21/42, 51/84, 29/41 \\
NLL $\alpha_S(\mu)$BPT & 114,82,67 & 1.29,1.49,1.62 & 19/42,64/84,37/41 \\
\hline
\end{tabular}
Table 2.  Sensitivity of our fit values for $\Lambda_{\overline{MS}}$ and $C_1$ to the choice
of fit range for heavy jet mass.  The fit ranges are $0.05M_Z/Q-0.12$ (low), $0.05M_Z/Q-0.24$ (normal), $0.125M_Z/Q-0.24$ (high).
\end{table}

\begin{figure}
\begin{center}
\includegraphics[scale=0.5]{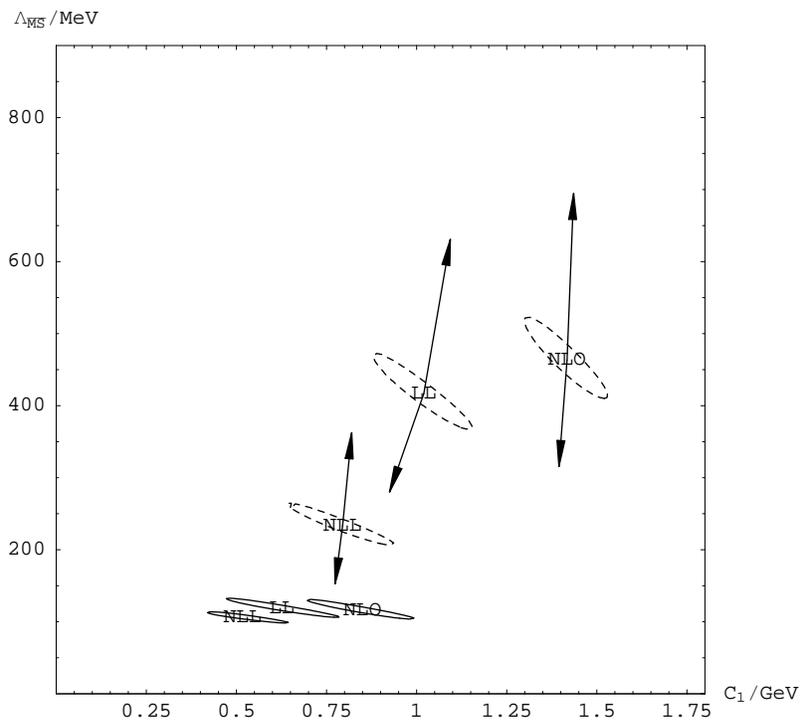}
\end{center}
\caption{Fits to the thrust distribution for $\Lambda_{\MSb}$ and $C_1$.  Solid ellipses use $\Lambda$BPT, dashed ellipses $\alpha_s(\mu)$BPT
(with the arrows showing the effect on the central value of varying $Q/2<\mu<2Q$). The ellipses indicate 2$\sigma$ errors generated by allowing
$\chi^2$ to vary within 4 of its minimum.  For a summary of the data used see Table 3. }
\label{fg:thrust}
\end{figure}

\begin{figure}
\begin{center}
\includegraphics[scale=0.5]{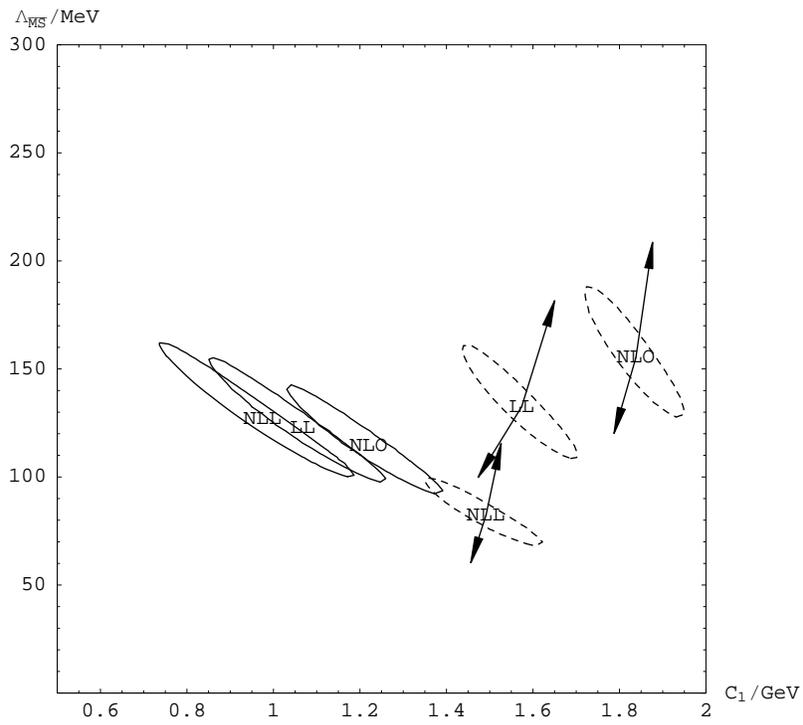}
\end{center}
\caption{As \reffg{thrust} but for heavy jet mass.  For a summary of the data see Table 4. }
\label{fg:hjm}
\end{figure}

\begin{figure}
\begin{center}
\includegraphics[scale=0.5]{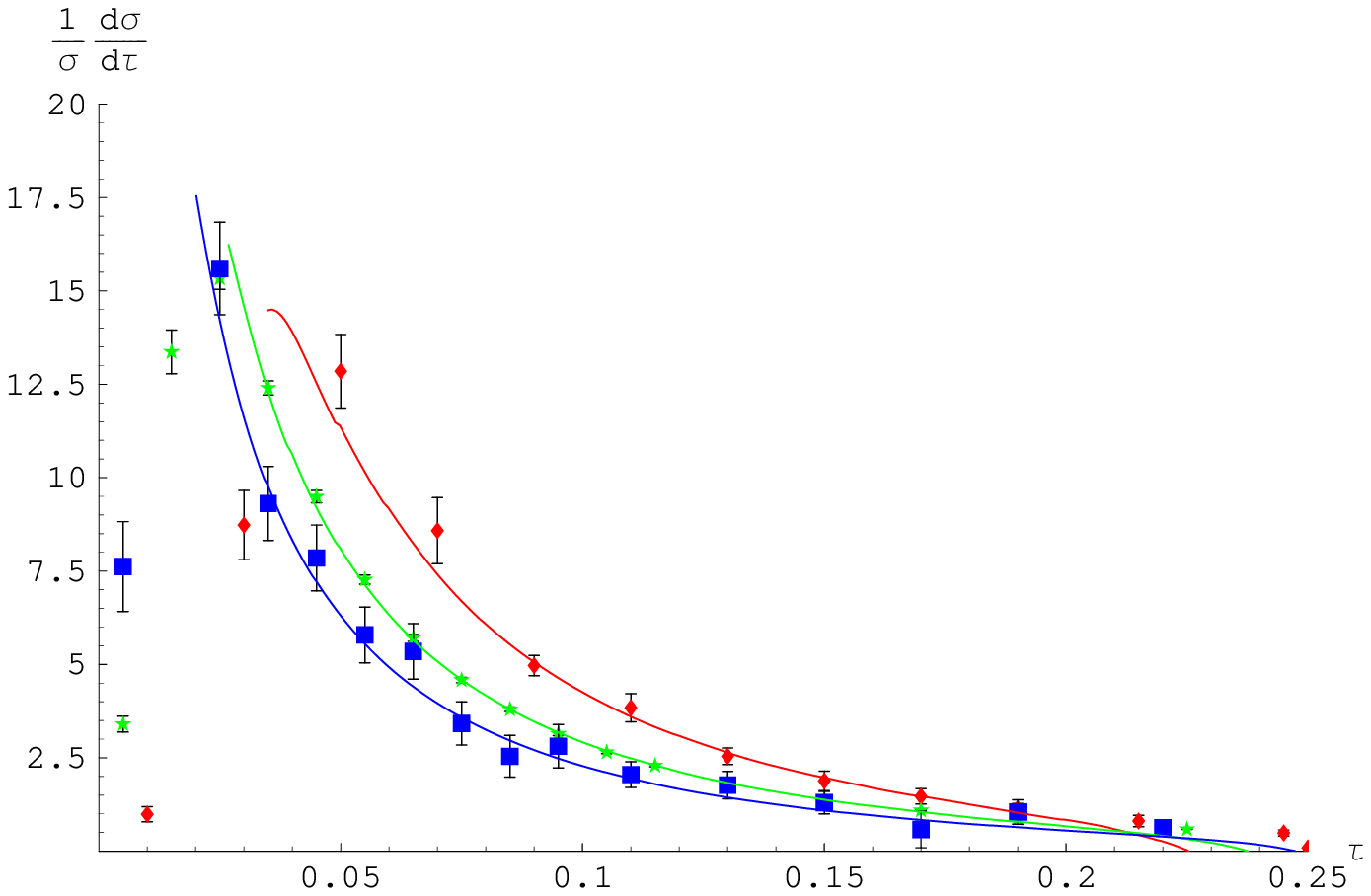}
\end{center}
\caption{Examples of our best fit NLO $\Lambda$BPT 1-thrust distributions. Red is JADE data at 44GeV, green is DELPHI at 91.2GeV and blue is DELPHI at 183GeV.}
\label{fg:nloFits}
\end{figure}

\begin{figure}
\begin{center}
\includegraphics[scale=0.5]{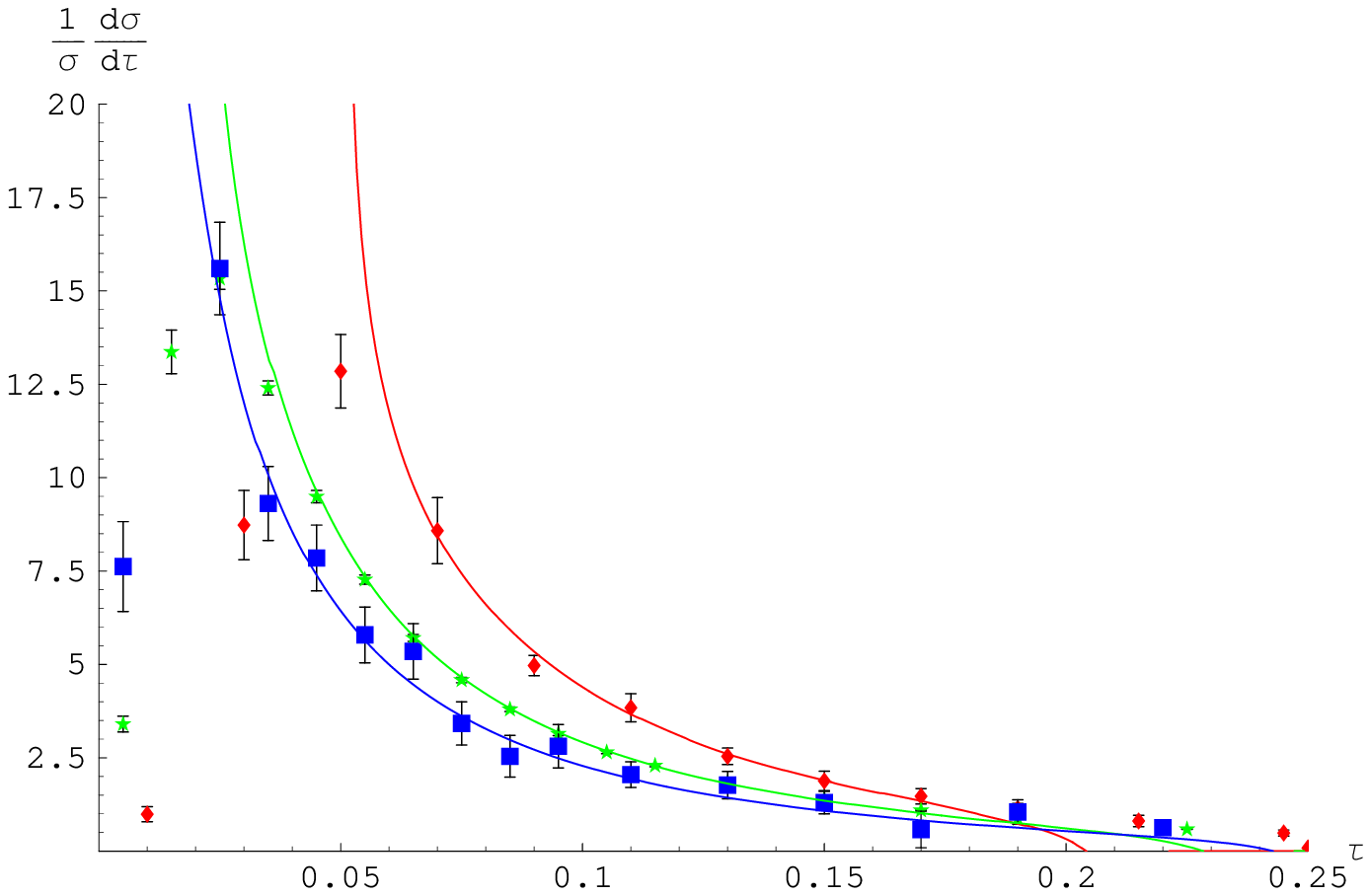}
\end{center}
\caption{Examples of our best fit NLL $\Lambda$BPT 1-thrust distributions. Red is JADE data at 44GeV, green is DELPHI at 91.2GeV and blue is DELPHI at 183GeV.}
\label{fg:nllFits}
\end{figure}

\begin{table}
\begin{tabular}{|l|c|c|c|c|}
\hline
Experiment & Q & Range & Data Points & Source\\
\hline
ALEPH & 91.2 & 0.05-0.18 & 7 & \cite{Buskulic:1992hq}\\
 & 133 & 0.04-0.15 & 4 & \cite{Buskulic:1996tt}\\
\hline
DELPHI & 91.2 & 0.05-0.18 & 10& \cite{Abreu:2000ck} \\
 & 133 & 0.04-0.18 & 5 & \cite{Abreu:1999rc}\\
 & 161 & 0.04-0.18 & 5 & \cite{Abreu:1999rc}\\
 & 172 & 0.04-0.18 & 5 & \cite{Abreu:1999rc}\\
 & 183 & 0.03-0.18 & 11 & \cite{Abreu:1999rc}\\
\hline
JADE & 35 & 0.14-0.18 & 2 & \cite{MovillaFernandez:1997fr}\\
 & 44 & 0.12-0.18 & 3 & \cite{MovillaFernandez:1997fr}\\
\hline
L3 & 91.2 & 0.065-0.175 & 4 & \cite{Adeva:1992gv}\\
 & 189 & 0.025-0.175 & 6 & \cite{Acciarri:2000hm}\\
\hline
OPAL & 161 & 0.03-0.15 & 6 & \cite{Ackerstaff:1997kk}\\
 & 172 & 0.03-0.15 & 6 & \cite{Abbiendi:1999sx}\\
 & 183 & 0.03-0.15 & 6 & \cite{Abbiendi:1999sx}\\
 & 189 & 0.03-0.15 & 6 & \cite{Abbiendi:1999sx}\\
\hline
SLD & 91.2 & 0.06-0.16 & 3 & \cite{Abe:1994mf}\\
\hline
TASSO & 44 & 0.12-0.16 & 1 & \cite{Braunschweig:1990yd}\\
\hline
\end{tabular}\\
Table 3. Summary of the data used in our fits for thrust.
\end{table}

\begin{table}
\begin{tabular}{|l|c|c|c|c|}
\hline
Experiment & Q & Range & Data Points & Source \\
\hline
DELPHI & 91.2 & 0.05-0.2 & 8 & \cite{Abreu:2000ck}\\
 & 161 & 0.04-0.2 & 4 & \cite{Abreu:1999rc}\\
 & 133 & 0.04-0.2 & 4 & \cite{Abreu:1999rc}\\
 & 172 & 0.04-0.2 & 4 & \cite{Abreu:1999rc}\\
 & 183 & 0.03-0.24 & 10 & \cite{Abreu:1999rc}\\
\hline
SLD & 91.2 & 0.08-0.24 & 3 & \cite{Abe:1994mf}\\
\hline
ALEPH & 91.2 & 0.05-0.2 & 8 & \cite{Buskulic:1992hq}\\
\hline
L3 & 91.2 & 0.051-0.216 & 7 & \cite{Adeva:1992gv}\\
 & 189 & 0.03-0.24 & 14 & \cite{Acciarri:2000hm} \\
\hline
OPAL  & 91.2 & 0.0625-0.2025 & 4 & \cite{Acton:1992fa}\\
 & 161 & 0.0289-0.2025 & 5 & \cite{Ackerstaff:1997kk} \\
 & 172 & 0.0289-0.2025 & 5 & \cite{Abbiendi:1999sx} \\
 & 183 & 0.0289-0.2025 & 5 & \cite{Abbiendi:1999sx} \\
 & 189 & 0.0289-0.2025 & 5 & \cite{Abbiendi:1999sx}\\
\hline
\end{tabular}\\
Table 4. Summary of the data used in our fits for heavy jet mass.
\end{table}

\section*{6 Discussion and Conclusions}
In this paper we have advocated the idea of $\Lambda$BPT in which
QCD observables ${\cal{R}}$ are directly related to the dimensional
transmutation parameter of the theory. Dimensional analysis
implies that this relation, given in Eq.(2), holds in general,
but explicit construction of the form of the function ${\tilde{\Phi}}({\cal{R}},\{{m}_{j}\}/Q)$
has only proved possible when ${\cal{R}}$ is simply related to an
effective charge. The testing of perturbative QCD then reduces to
direct extraction of the universal parameter ${\Lambda}_{\overline{MS}}$
from the measured experimental data
for the observable ${\cal{R}}$ using the fullest information available
about the function ${\tilde{\Phi}}$. For effective charges and with massless quarks we have
shown that this function has the structure ${\cal{K}}_{\cal{R}}^{\overline{MS}}{\cal{F}}({\cal{R}}){\cal{G}}({\cal{R}})$,
as in Eq.(24). Here the normalisation constant
${\cal{K}}_{\cal{R}}^{\overline{MS}}$ is obtainable from a NLO perturbative
calculation, and is the only part to depend on the subtraction convention used
to remove UV divergences. ${\cal{F}}({\cal{R}})$
is a universal function given in
Eq.(11), and if we only have perturbative information our knowledge of
${\cal{G}}({\cal{R}})$ is restricted to the perturbation series in Eq.(12),
where the coefficients ${g}_{i}$ (Eq.(13)) are RS-invariant combinations
of ${\alpha}_{s}({\mu})$BPT perturbative coefficients ${r}_{i}$ and beta-function
coefficients $c_i$. Given a ${\rm{N}}^{n}$LO calculation the first $n-1$
coefficients ${g}_{1},{g}_{2},\ldots,{g}_{n-1}$ are known. So at NLO all
we know about ${\cal{G}}({\cal{R}})$ is ${\cal{G}}({\cal{R}})=1$. As we
showed in Section 3 standard RG-improved ${\alpha}_{s}(\mu)$BPT is
exactly equivalent to ${\Lambda}$BPT with a particular RS-dependent
choice for the unknown higher-order RS-invariants ${g}_{i}\;(i\geq{n})$.
With the ``physical'' scale choice we showed that these unknown
coefficients can be extremely large, and that therefore the value
of ${\Lambda}_{\overline{MS}}$ obtained is potentially unreliable. It
seemed to us clearly more sensible to simply use only the known terms
and truncate Eq.(12) after $n-1$ terms at ${\rm{N}}^{n}$LO.\\

We should note that this dimensional analysis motivation was used by
Grunberg in Ref.\cite{r7} to motivate his method of effective charges.
Although slightly different in motivation the RESIPE approach of
Dhar and Gupta in Ref.\cite{r10} is exactly equivalent to Grunberg's
approach. In the method of effective charges one would truncate
the effective charge beta-function $B(x)$ in the integrand of $G({\cal{R}})$
in Eq.(16) setting $B(x)=1+cx+{\rho}_{2}{x}^{2}+\ldots+{\rho}_{n}{\cal{R}}^{n}$
given a ${\rm{N}}^{n}$LO calculation. On exponentiating to obtain ${\cal{G}}({\cal{R}})$
one would then partially include contributions $O({\cal{R}}^{n})$ and higher, so
the resulting ${\cal{G}}({\cal{R}})$ would differ by $O({\cal{R}}^{n})$ compared to
the strict truncation ${{\cal{G}}}^{(n)}({\cal{R}})$. The ${\Lambda}$BPT
designation is intended to provide a unifying motivation for these
various approaches, and to reinforce the fundamental importance
of the dimensional transmutation parameter.\\

In Section 4 we considered how Eq.(24) obtained with massless quarks is
modified if the quarks have masses. With massive quarks one would normally
employ Weinberg's form of the RG equation in terms of a running quark mass
\cite{r18}. In contrast we have started from the RESIPE formalism results
of Gupta, Shirkov and Tarasov \cite{r17} in which the physical pole mass
is used, these are equivalent to the
Lie differential RG equation developed
by Bogoliubov and Shirkov \cite{r19}. Our key result is that in
mass-independent schemes such as ${\overline{MS}}$ the ${\rho}_{i}$ RS-invariant combinations
of pertubative and beta-function coefficients are still RS-invariants,
which now depend on $\{{m}_{j}\}/Q$. We were able to prove that the
${\Lambda}$BPT result of Eq.(24) also holds in the massive quark case,
with $\{{m}_{j}\}/Q$-dependent coefficients ${g}_{i}$, given by exactly the
same combinations of ${\rho}_{i}$ as in the massless case (Eqs.(13)).\\

In Section 5 we extended the direct extraction of ${\Lambda}_{\overline{MS}}$ from
${e}^{+}{e}^{-}$ event shape observables of Ref.\cite{r9} to include a resummation of
large infra-red logarithms $L={\log}(1/y)$. One could relate the observable
${R}_{y}({y}^{\prime})$ to an effective charge ${\cal{R}}$ by
exponentiation, ${R}_{y}({y}^{\prime})={\exp}({r}_{0}{\cal{R}})$. One could then numerically
construct ${\rho}_{LL}({\cal{R}})$ and ${\rho}_{NLL}({\cal{R}})$ functions
by resumming to all-orders the corresponding pieces of ${\rho}_{n}$ in Eq.(69). The LL and NLL
predictions for ${R}_{y}({y}^{\prime})$ for a given value of ${\Lambda}_{\overline{MS}}$ then follow on inserting these
${\rho}({\cal{R}})$ functions in Eq.(16) and numerically solving the transcendental equation.
To model $1/Q$ power corrections we fitted to a shifted distribution ${R}_{PC}(y)=R_{PT}(y-{C}_{1}/Q)$.
Whilst in principle straightforward a number of complications arose. In particular as $1-T$ approaches
$1/3$ the leading coefficient $r_0$ goes to zero, invalidating the effective charge approach. This places
a rather stringent upper limit on the fit range.
There are also problems in the two-jet region as ${\Lambda}_{\cal{R}}$ approaches $Q$, which determine
the lower limit. We also noted that one cannot directly relate the observables to an unexponentiated
effective charge, as in Ref.\cite{r9}, since in that case ${\rho}_{LL}({\cal{R}})$ has a branch cut, so that ${\cal{R}}$ becomes undefined.
Simultaneous fits for ${\Lambda}_{\overline{MS}}$ and ${C}_{1}$ were performed using data
for thrust and heavy jet mass distributions over a wide range of energies (see Tables 3 and 4).
The $2{\sigma}$
error contours in ${\Lambda}_{\overline{MS}}$ and ${C}_{1}$ are shown in Figs.3 and 4. NLO,
LL and NLL results are shown for both standard ${\alpha}_{s}(\mu)$BPT with physical scale, and
for ${\Lambda}BPT$. For ${\alpha}_{s}(\mu)$BPT there is a strong decrease in ${\Lambda}_{\overline{MS}}$
going from NLO to LL to NLL, whereas for ${\Lambda}BPT$ the fitted ${\Lambda}_{\overline{MS}}$ values
are remarkably stable. The fitted value of ${C}_{1}$ is also somewhat smaller for ${\Lambda}BPT$.
We also investigated the stability of the fits to changing the fit range in Tables 1 and 2. The
${\Lambda}$BPT results show significantly more stability than ${\alpha}_{s}(\mu)$BPT.\\

We would conclude that, notwithstanding the limited fit range and technical complications from
which ${\Lambda}$BPT suffers, there is evidence that the fitted power corrections are reduced
relative to the standard approach, although not as dramatically as in the DELPHI fits of
Ref.\cite{r12} which are consistent with zero power corrections. However in that analysis
corrections for bottom quark mass effects were made, which were not included in our analysis.
Event shape means have also been measured in DIS at HERA \cite{Herameans}, and it would be interesting
to perform a $\Lambda$BPT analysis in that case as well. Unfortunately, as we remarked earlier,
it is not known how to construct the ${\tilde{\Phi}}$ function in Eq.(2) for DIS where one has
a convolution of parton distributions and hard scattering QCD cross sections. One can, however,
apply PMS to choose the renormalisation and factorisation scales. An analysis along these lines
is in progress \cite{mjd}.

\section*{Acknowledgements}
M.J.D. gratefully acknowledges receipt of a PPARC UK Studentship.

\end{document}